\documentclass[journal]{IEEEtran}

\usepackage[numbers,sort&compress]{natbib}

\usepackage[utf8]{inputenc}
\usepackage[T1]{fontenc}
\usepackage{amsmath}
\usepackage{algorithmic}
\usepackage{array}
\usepackage{siunitx}
\usepackage[caption=false,font=normalsize,labelfont=sf,textfont=sf]{subfig}
\usepackage{graphicx}
\usepackage{subfig}
\usepackage{multirow}
\usepackage{tabularray}
\usepackage{url}
\usepackage{textcomp}
\usepackage{bm}
\usepackage{float}

\begin{document}

\title{ExposNet: A Deep Learning Framework for EMF Exposure Prediction in Complex Urban Environments}

\author{Yarui~Zhang,
        Shanshan~Wang,~\IEEEmembership{Member,~IEEE,}
        and~Joe~Wiart,~\IEEEmembership{Senior Member, IEEE}

\thanks{Yarui Zhang, Shanshan Wang and Joe Wiart are with the Chaire C2M, Information Processing and Communications Laboratory (LTCI), Télécom Paris, Institut Polytechnique de Paris, 91120 Palaiseau, France (e-mail:~yarzhang@telecom-paris.fr; shanshan.wang@telecom-paris.fr; joe.wiart@telecom-paris.fr).}}

\markboth{Preprint}%
{Shell \MakeLowercase{\textit{et al.}}: Bare Demo of IEEEtran.cls for IEEE Journals}

\maketitle

\begin{abstract}
The prediction of the electric field (E-field) plays a crucial role in monitoring radiofrequency electromagnetic field (RF-EMF) exposure induced by cellular networks. In this paper, a deep learning framework is proposed to predict E-field levels in complex urban environments. First, the measurement campaign and publicly accessible databases used to construct the training dataset are introduced, with a detailed explanation provided on how these datasets are formulated and integrated to enhance their suitability for Convolutional Neural Networks (CNNs)-based models. Then, the proposed model, ExposNet, is presented, and its network architecture and workflow are thoroughly explained. Two variations of the network structure are proposed, and extensive experimental analyses are conducted, demonstrating that ExposNet achieves good prediction accuracy with both configurations. Furthermore, the generalization capability of the model is evaluated. The overall results indicate that, despite being trained and tested on real-world measurements, the model performs well and achieves better accuracy compared to previous studies.
\end{abstract}


\IEEEpeerreviewmaketitle

\section{Introduction}

\IEEEPARstart{T}{oday}, wireless communication relying on radiofrequency electromagnetic fields (RF-EMF) has become increasingly widespread. As reported by \cite{GSMA}, there were more than $12.172$ billion mobile connections globally in $2025$, exceeding the total population by $4.084$ billion. This highlights the extensive adoption of cellular communication systems. 

Guidelines for protection, such as those issued by the International Commission on Non-Ionizing Radiation Protection \cite{ICNIRP2020}, and those issued by IEEE \cite{8859679}, are in place. Additionally, standardized procedures have been established and implemented to ensure compliance with these safety measures \cite{IEC_62232, wiart2016radio, ANFR_protocol, CenelecTC106x}. Despite these regulations, concerns about potential risks from RF-EMF exposure persist \cite{EUBARO}. This context emphasizes the importance of RF-EMF exposure monitoring that has been assigned to the objectives of the EU call "HORIZON-HLTH-2021-ENVHLTH-02-01" \cite{Call_Horizon}. Following this EU call, several projects have been funded (SEAWave, GOLIAT, ETAIN, and NextGEMS). The SEAWave project, \cite{SEAWAVE} and under which the research for this paper was conducted, has been selected in response to this call and aims to address EMF monitoring and related concerns.

Over the past several decades, significant efforts have been devoted to assessing, standardizing, and monitoring EMF exposure across different environments in Europe \cite{JALILIAN2019108517}. For instance, \cite{gajvsek2015electromagnetic} provided a large-scale assessment of RF exposure across Europe, highlighting regional variations and regulatory compliance. Similarly, in \cite{takovidis22}, the authors utilized publicly available data to monitor EMF exposure trends across multiple European cities. 

More specifically, RF-EMF exposure has been extensively analyzed in different real-world scenarios, e.g., urban environments, public transport systems, and indoor settings. \cite{Wassim_Gol_2023} analyzed RF-EMF exposure induced by base stations in different urban scenarios. \cite{School2023} and \cite{Serbia2024} provides insights into exposure variation across different school settings. \cite{ShoppingMall2021} examines RF-EMF exposure in high-density public shopping malls. It highlights significant spatial variability in exposure levels, influenced by factors such as node density, architectural layout, and wireless network distribution. In \cite{Residential2021}, the exposure level of residential areas near mobile phone base stations in France is evaluated. Besides these indoor environments, public transportation systems including buses, trains, and subways have also been extensively studied, as passengers are continuously exposed to signals from base stations and mobile devices. In \cite{Loizeau23}, the authors conducted a comparative study of ambient RF-EMF levels in outdoor and public transport environments in Switzerland over a seven-year period, providing insights into temporal exposure variations.  \cite{celaya19} explored personal exposure levels in public buses, while \cite{aerts15} assessed RF-EMF exposure from small cell deployments inside trains. Besides, \cite{zhang2024statistical} provides a comprehensive analysis employing statistical moments models on RF-EMF exposure levels in public transport systems of Paris. Additionally, in \cite{Yang24}, the authors integrated numerical and experimental methods to assess RF-EMF exposure in vehicle-to-everything communication environments.

With the widespread deployment of 5G networks, researchers have increasingly focused on exposure levels associated with new frequency bands, beamforming technologies, and small cell deployments \cite{5Grisk}\cite{Review5G2024}. \cite{xu21} analyzed maximum exposure levels from 5G multi-cell base station antennas, providing insights into compliance assessments. Besides, \cite{sakovic21} compared exposure levels between microcell and macrocell topologies using operational network measurements, emphasizing the trade-off between coverage and exposure in different deployment scenarios. As 5G networks introduce millimeter-wave mmWave frequencies, \cite{Narayanan} conducted a comparative measurement study on commercial 5G mmWave deployments, evaluating the real-world performance and exposure characteristics. \cite{Jiang2024} evaluates the RF-EMF exposure induced by wireless cellular phones under various usage scenarios in France, considering different network generations, including 3G, LTE, and 5G non-standalone (a.k.a. 5G NSA), providing insights into user-specific exposure based on measurement-based assessments.

To accurately assess RF-EMF exposure, a range of measurement techniques has been employed, including drive tests, sensor networks, and spot measurements. Many countries have deployed EMF probes \cite{sondeFR}\cite{sondeSerbia}\cite{Ourouk2024} to monitor temporal variations in electric field (E-field) strength, enabling long-term exposure assessments. However, these probes are restricted to fixed locations, limiting their ability to capture broader spatial variations.
Drive tests, conducted using vehicles equipped with spectrum analyzers, provide spatially distributed measurements along predefined routes. This method effectively captures spatial variations in exposure levels within a short timeframe (typically several hours). In \cite{SW}, a hybrid approach integrating drive test data with sensor network measurements was used to monitor RF-EMF exposure in a French city, demonstrating the advantages of combining different measurement strategies for a more comprehensive assessment.
Additionally, spot measurements \cite{ANFR_carto}\cite{ANFR_mesures}\cite{sondeSerbia}, a form of static measurement using spectrum analyzers fixed on tripods, offer highly localized, precise exposure evaluations. These measurements are often conducted in response to public and regulatory concerns, ensuring targeted assessment in areas of interest.
Each of these measurement approaches presents trade-offs in accuracy, cost, and feasibility. Sensor networks allow for continuous monitoring but require extensive infrastructure deployment. Drive tests offer broad spatial coverage but are limited in temporal resolution. Spot measurements provide high accuracy yet only capture short-term snapshots. Finally, the challenge remains that comprehensive large-scale measurements are both time-consuming and costly, making it impractical to cover all locations extensively.

To complement direct RF-EMF exposure measurements, various modeling approaches have been developed to estimate exposure levels efficiently across diverse environment. Stochastic geometry is usually used to derive statistical exposure distributions for large-scale networks but lacks the ability to provide precise, location-specific estimates \cite{SG1}\cite{Gontier21}. Kriging interpolation is useful for the interpolation of exposure levels between sparse measurement points, producing smooth spatial maps. However, its accuracy heavily depends on the density of input data and may struggle in highly heterogeneous environments \cite{Kriging2022}\cite{Lemaire2016}. 
As for ray tracing simulations \cite{UrbanRayTracing}, it can offer high-fidelity exposure predictions by accounting for environmental interactions such as reflections and diffractions, but it requires extensive computational resources and detailed environmental datasets, limiting their scalability.

To address the limitations of traditional modeling approaches, machine learning (ML)-based methods have emerged as a powerful alternative for EMF exposure prediction, offering the ability to efficiently model complex relationships between multiple influencing factors. ML techniques have already been applied in various EMF exposure studies. For instance, graph neural networks (GNNs) have been utilized to predict EMF distribution in indoor environments, taking into account transmitter (Tx) positions and geometries, demonstrating strong generalization capabilities in unseen environments \cite{GNN_indoor_NICT}. In another study, a reconfigurable neural network architecture (RAWA-NN) was designed to predict absorbed power density (APD) and temperature rises in human tissue \cite{NNforHumanBody}. Moreover, an artificial neural network (ANN) was trained on measurement data from multi-floor buildings to estimate UL transmitted power in indoor scenarios \cite{ULANN}.

The objective of this contribution is to explore how ML-based methods can be leveraged for predicting EMF exposure levels in outdoor urban environments. Several studies have already tackled this issue. \cite{MallikCNN} proposed an infinitely wide convolutional neural network (CNN) model for EMF exposure reconstruction, providing a potential solution for reconstructing exposure maps from sparse sensor data. However, their model was trained with simulated targets, and in real-world applications, sensors cannot be installed across the entire city to achieve high-resolution exposure maps. Another approach proposed in \cite{IETBeijing} employed a random forest regression model to predict E-field intensity near 5G base stations, using factors such as transmit power, antenna gain, and environmental complexity collected from field measurements. Nevertheless, in practical applications, key data such as transmit power and antenna gain of base station antennas (BSAs) are not always publicly available. For example, in France, Cartoradio \cite{ANFR_carto} provides BSAs information, including location, support height, frequency bands, and azimuth, but lacks crucial parameters like antenna tilt, antenna gain, and transmit power.

To mitigate these data limitations, previous studies have explored how to combine real-life measurements with partially available public data to train ANN models for EMF exposure prediction in urban environments. \cite{SensorAidedANN} first investigated the feasibility of using ANNs trained on simulated data to predict exposure levels in urban outdoor settings. This work was later extended in \cite{telecom_Wang}, who conducted real-world drive test measurements and incorporated them into ANN training. \cite{ChikhaAccess} further refined this approach by introducing a feature selection strategy based on propagation models and the Gram-Schmidt Orthogonalization procedure to optimize the ANN input set.

This paper aims to develop a deep learning framework that leverages real-world measurement data and publicly accessible but partially incomplete BSAs information to predict EMF exposure levels in complex urban environments. Recent advancements have demonstrated the effectiveness of CNNs in path loss prediction \cite{PropagationCNN1, RadioUnet, Zhang2020CellularNR, ModelAided}, where transmitter and receiver locations, geographic information system (GIS) data, including terrain information, building structures, and antenna details, is formulated as two-dimensional ($2$-D) spatial representations. CNNs, which are powerful tools for extracting features from structured spatial data, have been successfully used to capture complex radio propagation patterns.

However, unlike path loss prediction, EMF exposure estimation involves the cumulative effect of multiple base station antennas, making it inherently more complex due to the aggregation of all received powers plus environmental noise. Some studies have attempted to adapt CNN-based approaches for EMF exposure modeling. For example, in \cite{5G_Unet_Korea}, a deep learning model based on a U-Net architecture was proposed to predict EMF exposure from 5G base stations, where GIS information and detailed antenna parameters were encoded as a $2$-D input matrix. However, this approach was also trained on simulation data and required comprehensive base station parameters, limiting its applicability in real-world scenarios.

Inspired by the advancement of CNNs in path loss prediction, the present contribution proposes a deep learning approach that relies solely on measurement data and publicly available datasets, without requiring exhaustive simulation inputs nor sophisticated base station parameters.
The main contributions are as follows:
\begin{itemize}
    \item Development of a novel data integration strategy that combines drive test measurements, BSAs database, and geographic information into a structured $2$-D input format suitable for CNN-based models. 
    \item Design of a flexible deep learning framework with two prediction modes: frequency-selective prediction and total E-field prediction. This framework enables the extraction of relevant input features and provides frequency-band-specific E-field predictions when frequency-selective datasets are available, while also supporting total E-field prediction when only broadband measurements are accessible.
    \item Development of a deep learning model with improved interpretability for predicting urban EMF exposure levels, while using real-world measurements for both training and evaluation to ensure reliability and practical applicability. In addition, measurements from $2$ cities are employed to improve the generalization ability of the trained model.
\end{itemize}


To the best of the authors' knowledge, this study is the first to propose a robust deep learning-based framework for EMF exposure prediction, utilizing $2$-D inputs derived from multi-modal data. Furthermore, real-world measurements from different cities are incorporated for both training and testing, ensuring the model's practical applicability and generalization. 

The rest of this paper is structured as follows. The problem statement is given in Section \ref{Section2}. The details about data preprocessing and organization is sketched in Section \ref{Section3}. The proposed learning framework is introduced in Section \ref{Section4}. The experimental setup is described in Section \ref{Section5}. Results and discussion are provided in Section \ref{Section6}. Conclusion and perspectives are in Section \ref{Section6}.
\section{Problem statement}\label{Section2}
Accurately predicting EMF exposure levels in urban environments is a critical task for network planning, public health assessment, and regulatory compliance. However, due to the interplay of multiple propagation factors, forecasting EMF exposure in such complex settings presents a significant challenge.

A fundamental issue lies in how to accurately model the path loss in dense urban areas, which is shaped by two primary factors: the structural characteristics of the city and the configuration of base stations. 
To capture the complexity of signal attenuation in complex urban settings, some propagation models such as the Walfisch-Ikegami model \cite{ikegami1991theoretical}, have been developed to account for additional characteristics specific to urban scenarios.
Despite the availability of propagation models tailored for urban settings, several challenges persist. 

First, urban environments exhibit high structural heterogeneity, with variations in building density, material properties, and spatial configuration. The signals may undergo diffraction at building edges, reflection from surfaces, and attenuation due to obstructions. The height of surrounding buildings, the width and orientation of streets, and the presence of moving vehicles and passengers all contribute to location-dependent signal attenuation. These factors introduce unpredictable propagation paths. As a result, EMF exposure levels can vary significantly within short distances, making accurate modeling particularly challenging.

Second, EMF exposure estimation extends beyond a simple path loss calculation, as it is influenced by the presence of multiple BSAs. The number, location, height, azimuth, tilt, and transmit power levels of these BSAs collectively contribute to fluctuations in EMF exposure. Additionally, the mobility of user equipment and dynamic traffic demand introduce further uncertainty, making exposure estimation even more complex.

Given this context, several key questions arise: 
\begin{enumerate}
    \item Real-world scenarios are highly complex and dynamic. In situations where we cannot obtain the complete information on the key factors influencing exposure levels, how to effectively integrate the available data?
    \item How to preprocess the on-site measurement data and select appropriate features to define a meaningful target for prediction?
    \item How to design an appropriate neural network model to process multi-modality input data, and predict EMF exposure levels?

\end{enumerate}

In the following sections,  we will introduce the datasets used in this study, related measurement campaigns and proposed deep learning framework while addressing these key questions in detail.
\section{Available datasets and measurement campaign}\label{Section3}

This section details the datasets used for the deep learning model, including publicly accessible base station dataset, geospatial database, and the on-site measurement conducted in urban environment.
\subsection{Base station and Geospatial dataset}
As discussed in Section \ref{Section2}, predicting EMF exposure levels in urban environments requires two essential types of information: base station data and geospatial environmental data, as both directly influence EMF propagation.  

In France, the National Frequency Agency (ANFR) manages a publicly accessible database Cartoradio \cite{ANFR_carto}, which contains detailed records of all registered BSAs with transmit power higher than \SI{5}{\watt} across the country. 

For each BSA, the available information includes their location, height, azimuth, type, as well as their operating frequency bands.
However, certain critical parameters are missing, such as the downtilt, and the emitted power of antennas.

The geospatial dataset used in this study is obtained from the National Institute of Geographic and Forest Information (IGN) \cite{IGN}. The extracted dataset contains the spatial distribution of buildings, i.e., building layouts, and height maps that provide information on building density within a given area. Additionally, satellite imagery, particularly in the infrared spectrum, is used to distinguish key surface features such as buildings, vegetation, and water bodies. Furthermore, land cover maps are incorporated to classify different surface types, offering a comprehensive understanding of the environmental context. By integrating these diverse geospatial datasets, we can derive a set of environmental features that influence EMF wave propagation for in urban scenarios.

\subsection{Drive test measurement in two French cities}
The drive test campaign was conducted in Paris and Lyon, two major metropolitan areas in France. The following parts of this section provide a detailed description of the measurement campaign, followed by an analysis of the collected data.
\subsubsection{Measurement equipment}
\begin{figure}[hbtp]
	\centering
	\subfloat[]{\includegraphics[width=.2\linewidth]{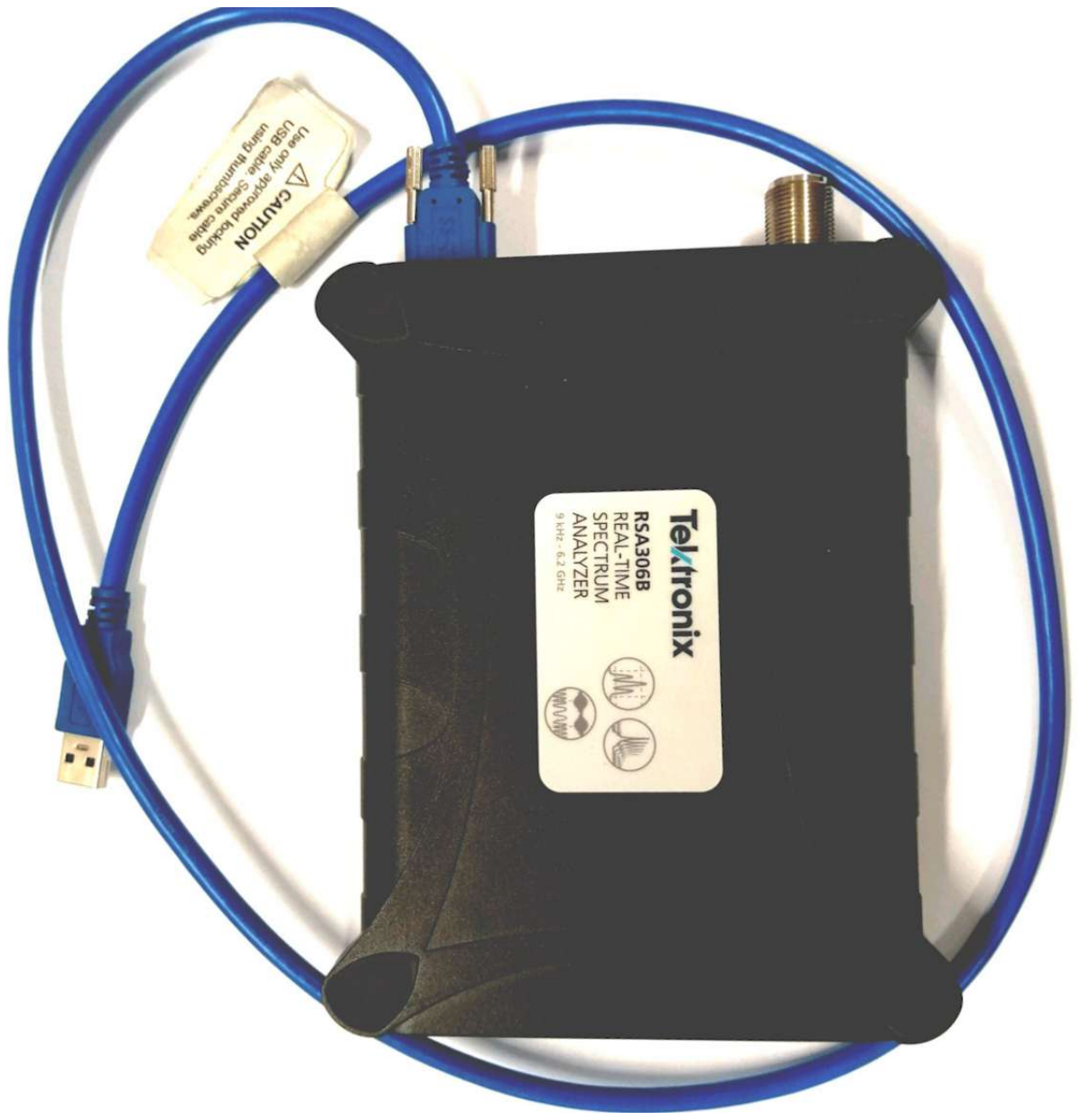}\label{Fig1a}} 
	\subfloat[]{\includegraphics[width=.2\linewidth]{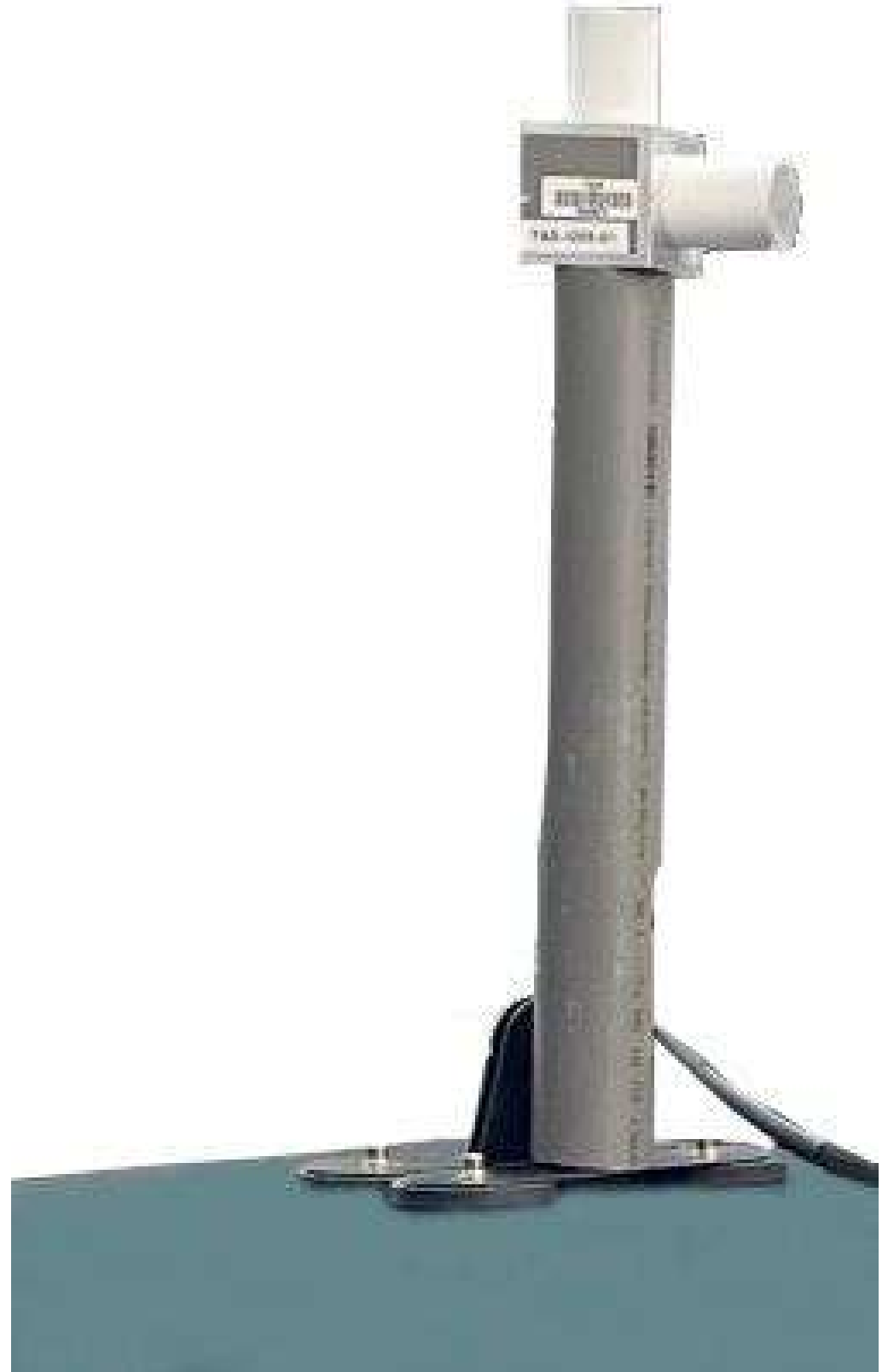}\label{Fig1b}}\hspace{3pt}
        \subfloat[]{\includegraphics[width=.2\linewidth]{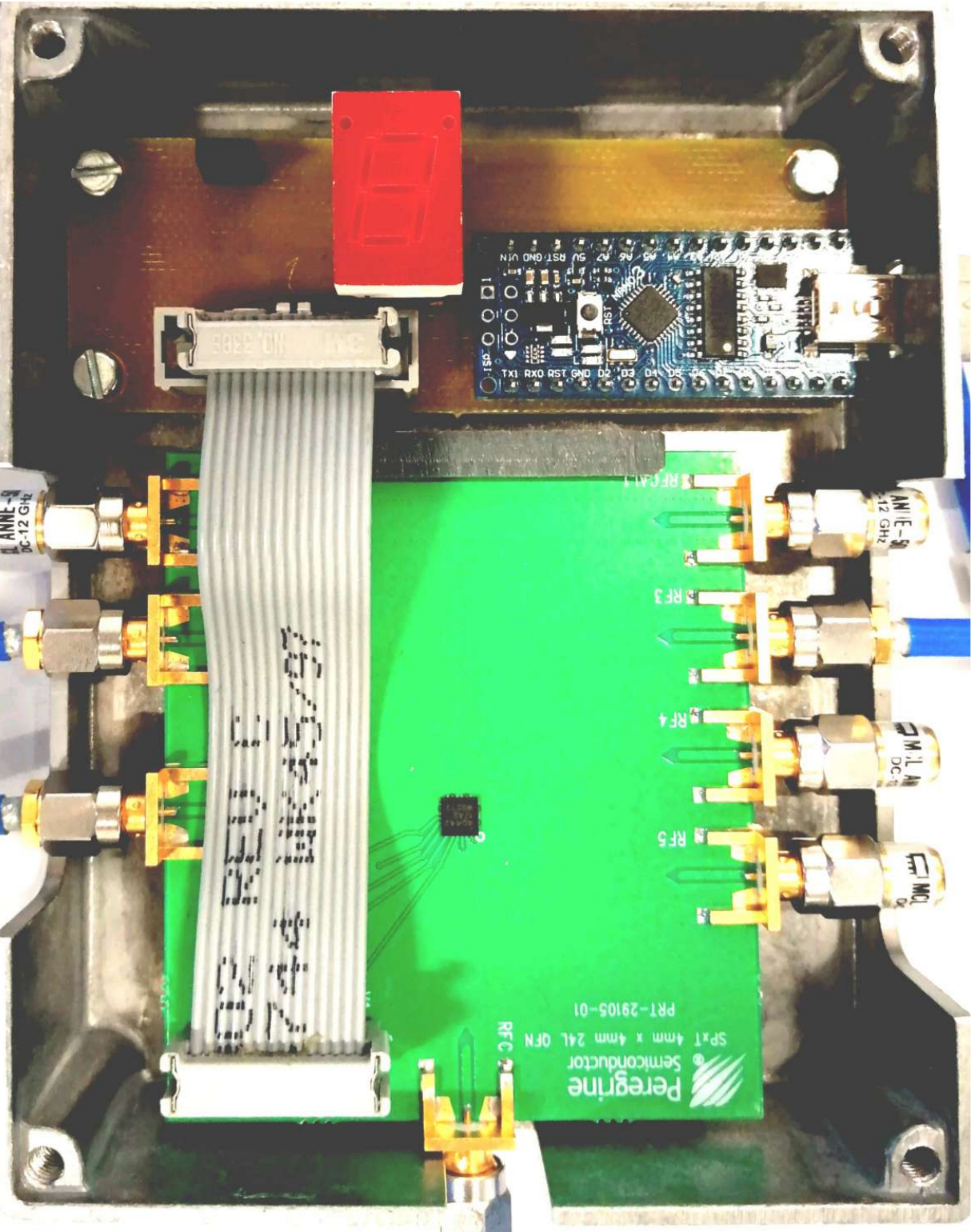}\label{Fig1c}} \hspace{3pt}
	\subfloat[]{\includegraphics[width=.2\linewidth]{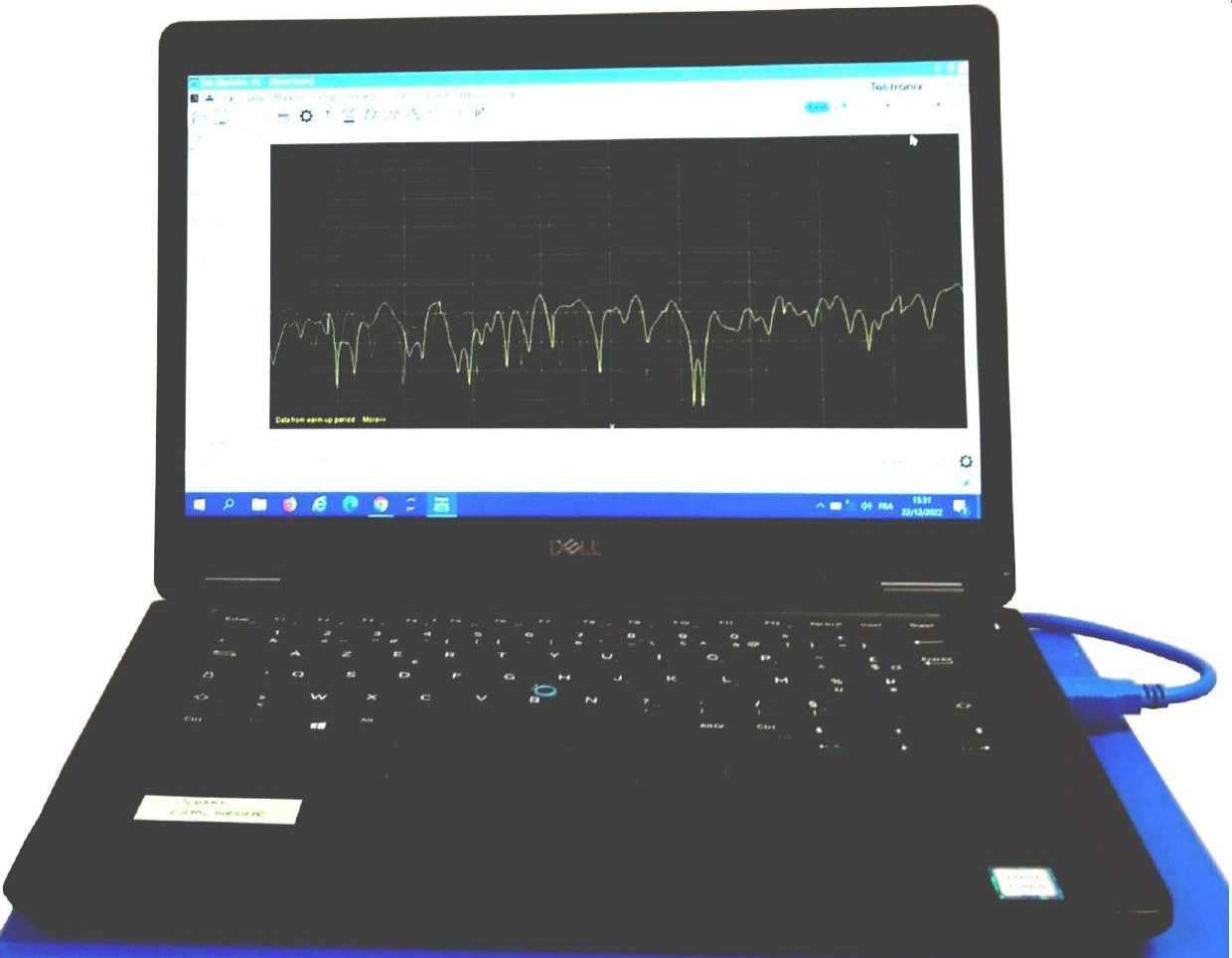}\label{Fig1d}}
\label{FigMesurementSystem}	
 \caption{Measurement system: \protect\subref{Fig1a} Tektronix RSA306B, \protect\subref{Fig1b} Tri-axis antenna probe, \protect\subref{Fig1c} Switch and \protect\subref{Fig1d} PC.  }
\end{figure}

The measurement campaign employed a measurement system consisting of a Tektronix RSA306B real-time spectrum analyzer (Fig.~\ref{Fig1a}), in conjunction with a TAS-1208-01 tri-axis antenna probe (Fig.~\ref{Fig1b}) comercialized by MVG \cite{MVG2025}, a switch (Fig.~\ref{Fig1c}) and a personal computer (PC) (Fig.~\ref{Fig1d}).

The RSA306B spectrum analyzer, known for its compact design and quick acquisition speed, was chosen for its ability to provide real-time spectrum analysis while being portable enough for outdoor mobile measurement campaigns. 

The antenna, capable of measuring E-field strength in three orthogonal polarizations (X, Y, and Z) from \SI{9}{\kilo\hertz} to \SI{6.2}{\giga\hertz}, was securely mounted on the roof of a vehicle to ensure stable data acquisition during movement. The tri-axis antenna was connected to the spectrum analyzer via an Arduino controlled switch, enabling automated sequential data collection from each axis without manual intervention. The isotropic E-field strength was computed using the equation: $E = \sqrt{\sum^3_{j=1}E^2_j}$, where $E_j$ represents the E-field measurement for each orthogonal axis. 
A custom-developed Python-based graphical user interface (GUI) was utilized to facilitate the configuration and control of the spectrum analyzer. This GUI allowed for the adjustment of key parameters, including resolution bandwidth (RBW), reference level, center frequency, and frequency span etc. 
In our setup, we consider the isotropic E-field measurements across a broad frequency range from \SI{700}{\mega\hertz} to \SI{3800}{\mega\hertz}, covering RF bands associated with 2G, 3G, 4G, and 5G networks.

\subsubsection{Drive test protocol}
\begin{figure}[hbtp]
	\centering
	\subfloat[]{\includegraphics[width=.98\linewidth]{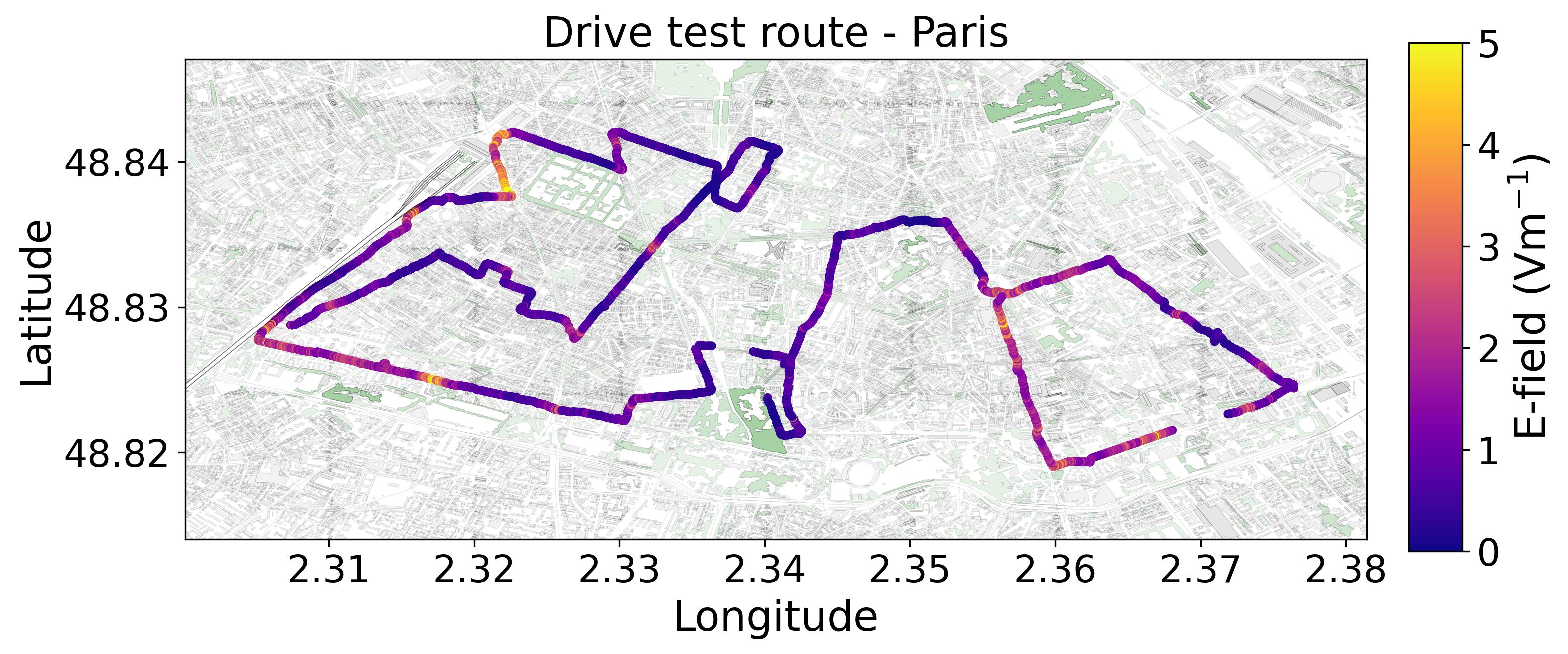}\label{Fig2a}} 
    
	\subfloat[]{\includegraphics[width=.98\linewidth]{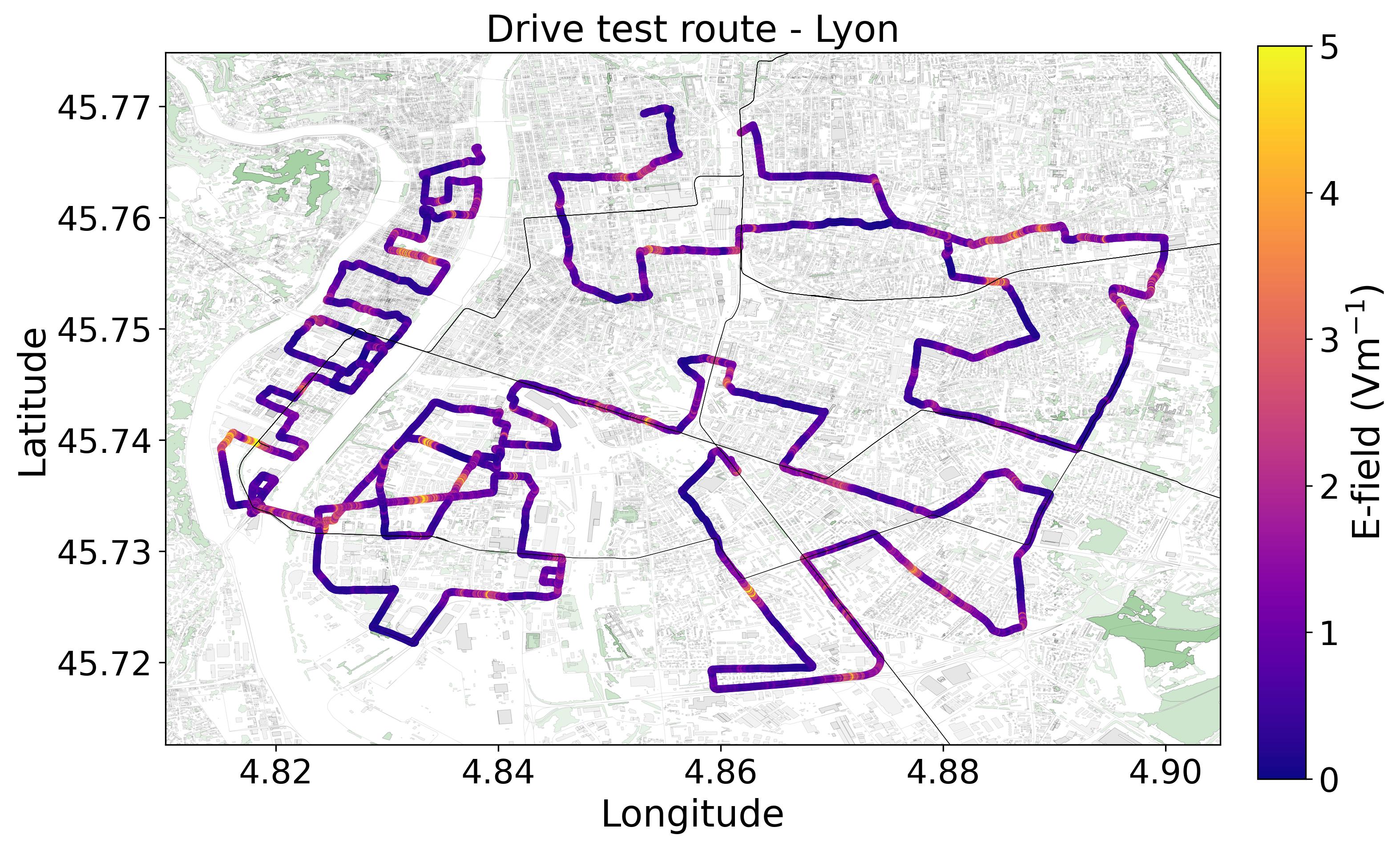}\label{Fig2b}}
\label{FigDriveTestRoute}	
 \caption{Drive test route: \protect\subref{Fig2a} Paris, \protect\subref{Fig2b} Lyon. }
\end{figure}

The drive test routes, illustrated in Fig.~\ref{FigDriveTestRoute}, was planned to encompass a diverse range of urban environments within Paris and Lyon, including residential areas, commercial districts, public facilities, and open spaces. The total route length was approximately \SI{22}{\kilo\meter} in Paris (Fig.~\ref{Fig2a}) and \SI{64}{\kilo\meter} in Lyon (Fig.~\ref{Fig2b}). The measurements were conducted over a single day during several hours for each city, mainly to capture variations in RF-EMF exposure due to spatial changes in environmental conditions.

To ensure measurement consistency and reduce the impact of vehicle motion on data accuracy, the vehicle maintained controlled speeds throughout the test, i.e., less than \SI{25}{\kilo\meter\per\hour}. The required time to record one complete $3$-axis measurement for the considered frequency range is around \SI{0.8}{\second}. The total number of measurements is $7516$ in Paris and $19062$ in Lyon, and the average distance between two successive measurement points is around \SI{2.76}{\meter} in Paris and \SI{3.36}{\meter} in Lyon.

Since the Tektronix RSA306B does not include an integrated GPS module, GPS coordinates for each measurement point were recorded using the ``Geo Tracker" mobile application and the GPS recording system from CSTB \cite{CSTB}. These GPS recorders provide timestamps and geographic coordinates (latitude and longitude), which will be synchronized with the E-field measurements to ensure accurate spatial referencing. 

\subsubsection{Descriptive analysis of measured data}
As previously mentioned, the measurements encompass a wide frequency range from \SI{700}{\mega\hertz} to \SI{3800}{\mega\hertz}, covering seven downlink frequency bands utilized by the four major French operators. Hereafter, we refer to these bands as \SI{700}{\mega\hertz}, \SI{800}{\mega\hertz}, \SI{900}{\mega\hertz}, \SI{1800}{\mega\hertz}, \SI{2100}{\mega\hertz}, \SI{2600}{\mega\hertz}, and \SI{3500}{\mega\hertz}. 

\begin{figure}
    \centering
    \includegraphics[width=0.9\linewidth]{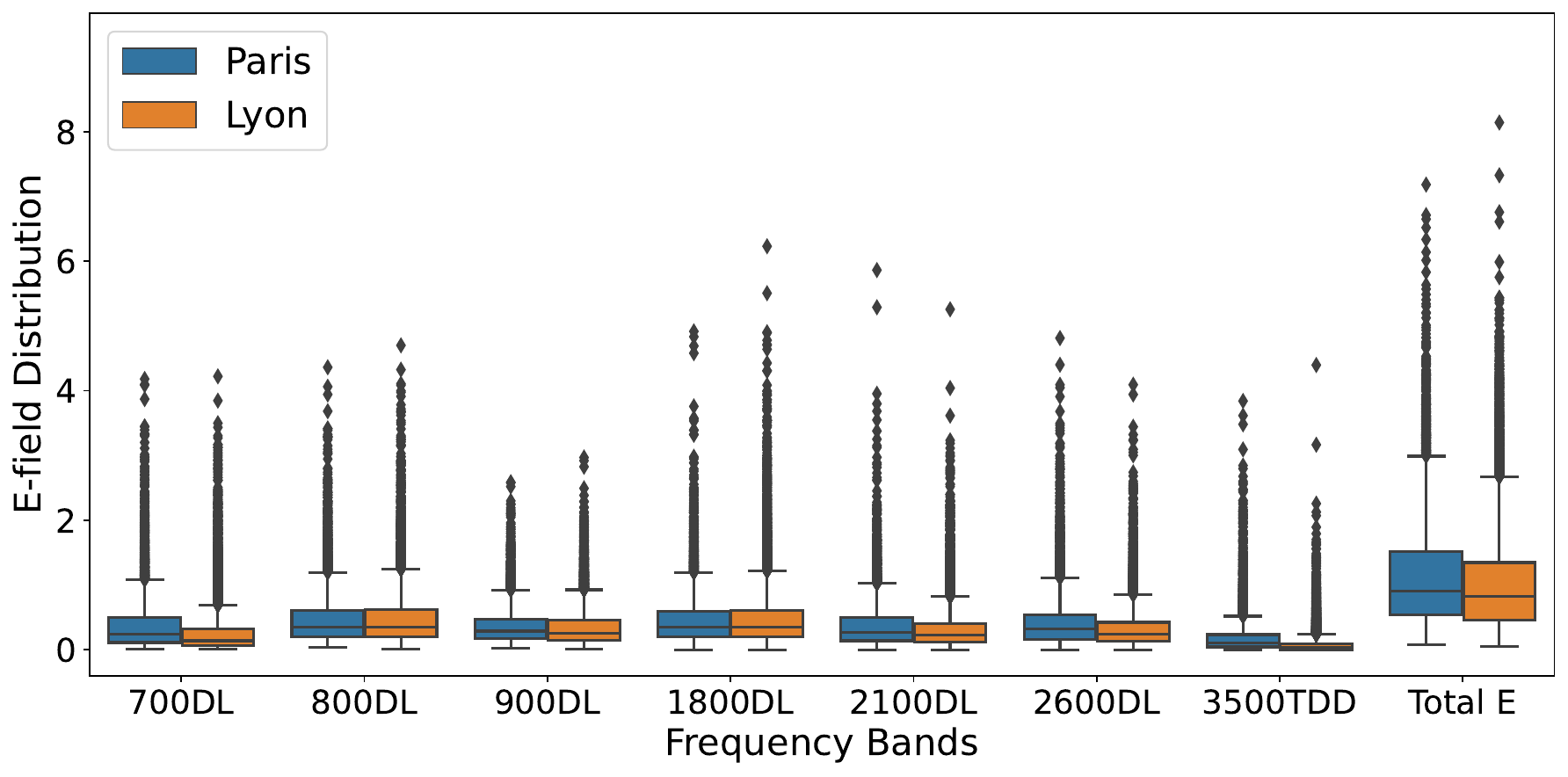}
    \caption{Box plots of measurement data in two cities. This plot shows the median (black line), interquartile range (box: $25$th–$75$th percentiles), and whiskers extending to $1.5\times$IQR. Outliers beyond this range are shown as black dots.}
    \label{fig:box_plot}
\end{figure}

Fig.~\ref{fig:box_plot} presents a comparative analysis of the distribution of the E-field values for each frequency band, along with the total E-field value, for the cities of Paris and Lyon. The total E-field is defined as: 
\begin{equation}
    E_{total} = \sqrt{\sum^{N_f}_{f=1} E_f^2}
    \label{EtotalCalculation}
\end{equation}

where $f$ denotes a specific frequency band, and $N_f$ represents the total number of frequency bands. Note that here the relative limits for different frequency bands are not taken into account. From the results, we can observe that the median and IQR in the \SI{800}{\mega\hertz} to \SI{2600}{\mega\hertz} bands are similar between Paris and Lyon, indicating that the primary measured data for both cities fall within comparable ranges. However, in the \SI{700}{\mega\hertz} and \SI{3500}{\mega\hertz} bands, the data distribution in Lyon is more concentrated, and the overall E-field is lower, which may be attributed to differences in 5G deployment between the two cities. In terms of total E-field, Paris exhibits a higher median and a slightly larger IQR, which suggests that the overall E-field value is larger compared to Lyon.

\section{Proposed deep learning scheme}\label{Section4}
In this section, a detailed explanation is provided on how the input and output of the proposed deep learning model are formulated. Subsequently, the originally designed deep neural network, ExposNet, is introduced. Its architecture, workflow, and the functionality of each component are thoroughly explained to provide a clear understanding of its overall structure.
\subsection{Input and output design}
As previously introduced, for any given area, we have access to a diverse set of data, including satellite imagery, building height maps, and nearby BSA parameters, sourced from publicly available databases.
To maximize the utility of this data, the input format must be carefully designed.  In this study, unlike previous work \cite{ChikhaAccess}, where the input features for the neural network were represented as single scalar values (e.g., the height of the nearest building or the distance from the measurement point to the closest building), the inputs are structured as a three-dimensional tensor of shape $C\times H \times W$. Here, $C$ denotes the number of channels, while $H$ and $W$ represent the height and width of the feature map, respectively. 

The key idea is to define a square area of $N \times N$ \SI{}{\meter\squared} centered on the GPS location of interest. Satellite imagery, building height map, and BSA maps for this selected area are then extracted and transformed into feature maps of dimensions $H \times W$, providing a richer and more spatially-informed input representation. Based on the above concept, the input tensor, as shown in Fig.~\ref{fig:Input}, is designed to have a total of $15$ channels. It is composed of the following elements:
\begin{figure}[htbp]
    \centering
    \includegraphics[width=0.9\linewidth]{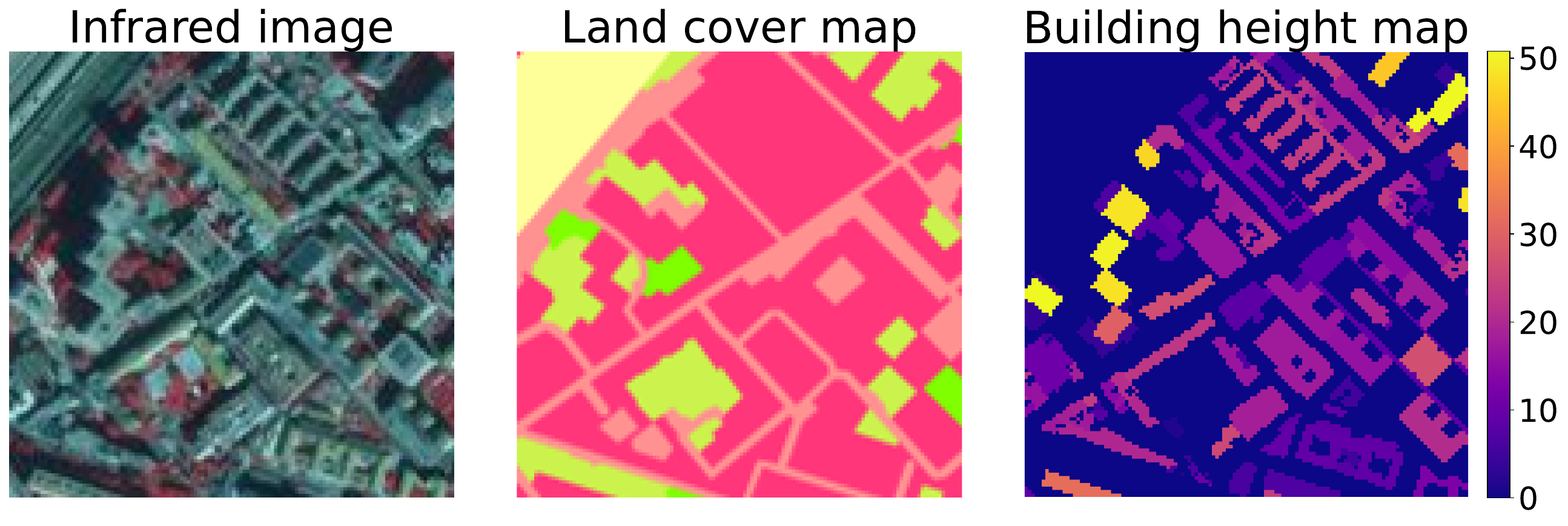}

    \includegraphics[width=0.91\linewidth]{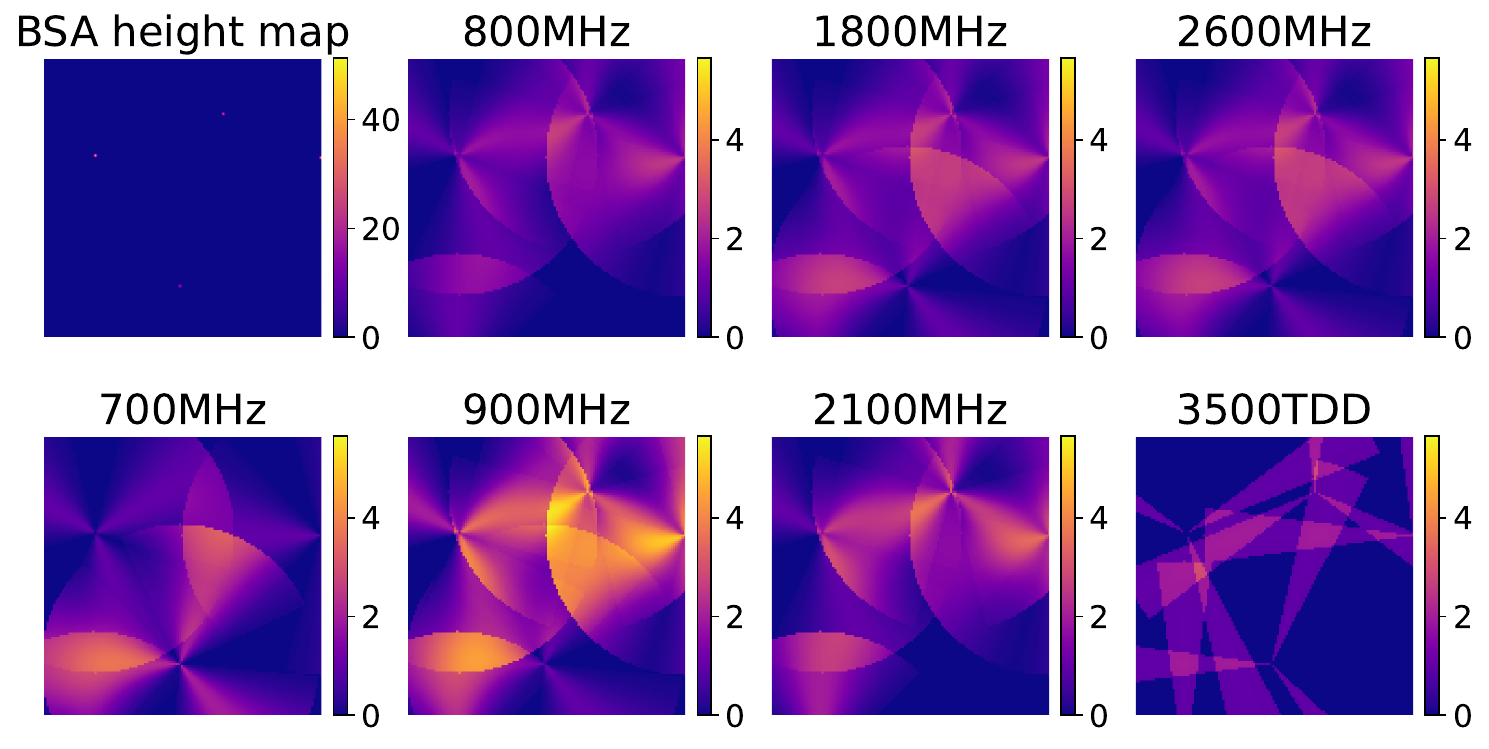}
    \caption{Example of channels in an input tensor. The unit of the color bar for the ``Building height map'' and ``BSA height map'' is meters. In the ``Antenna radiation maps'', the color bar values represent the superposition of normalized antenna gains.}
    \label{fig:Input}
\end{figure}

\begin{enumerate}
    \item Satellite infrared image ($3$ Channels):\\
    The infrared image captures information from the infrared spectrum, allowing the differentiation of various surface types such as water, vegetation, and buildings. Additionally, it provides a detailed representation of urban structures. As an RGB image, it consists of three channels, with each channel containing pixel values ranging from $0$ to $255$.
    \item Land cover map ($3$ Channels):\\
    The land cover map is also represented by $3$ channels since it is an RGB image. As illustrated in the 
    figure, unlike infrared images that capture detailed textures, the land cover map is composed of distinct color blocks, each representing a specific surface category. Examples include built-up areas, unbuilt areas, mineral surfaces, composite material surfaces, water surfaces, and various types of vegetation. Each surface type is encoded with a unique color, and the values for each channel range from $0$ to $255$. For locations where land cover information is unavailable, a fully white map (all channels set to $255$) is used as a placeholder. The land cover map complements the satellite infrared image, as it provides information on different land types within the given area, while more detailed building structures are only visible in the infrared image.
    
    \item Building height map ($1$ Channel):\\
    This input consists of a single channel that captures building layout and height information, which is not provided in the satellite image and land cover map. The pixel value at each location corresponds to the building height in meters. For locations without buildings, the pixel value is set to zero. If a building exists but height information is unavailable, the height is estimated based on an assumption of \SI{3}{\meter} per floor, provided the number of floors is known. If no floor information is available, the height is inferred from adjacent buildings. 
    \item BSA height map ($1$ Channel):\\
    This input represents the height map of base station sites within the area. Pixels corresponding to BSA locations are assigned a value equal to the average height of the antennas installed at the site. For all other pixels, where no base station is present, the value is set to $0$.
    \item Antenna radiation maps ($7$ Channels):\\
    This input represents the antenna radiation patterns within the area for seven frequency bands. Since the exact antenna gain is unknown, we assume the gain distribution follows $cos^8(\theta)$ model \cite{antenna_formula}, where the gain decreases as the angle deviates from the main lobe direction, with the maximum gain normalized to $1$. 
    While the exact beamwidth and downtilt angles are unavailable, information about the antenna type and installation scenario is utilized to assign reasonable estimates for the beamwidth and coverage range of each antenna. 
    Furthermore, the radiation pattern map is not restricted to antennas located within the target area; antennas located outside the area are also considered if their radiation patterns extend into the region of interest. This ensures that the spatial influence of all relevant antennas is captured in the input representation.
\end{enumerate}

With this $15$-channel input tensor, which contains rich environmental and antenna information for a given area, the prediction target is designed to include the root mean square (RMS) of E-field value across all measurement points within the area, as well as their standard deviation (STD). It is worth noting that the proposed deep learning model is highly flexible. By making minor adjustments to the network structure, it can accommodate both total E-field prediction and frequency-selective prediction, offering adaptability to various measurement scenarios.

\begin{figure*}[t]
    \centering
    \includegraphics[width=0.95\linewidth]{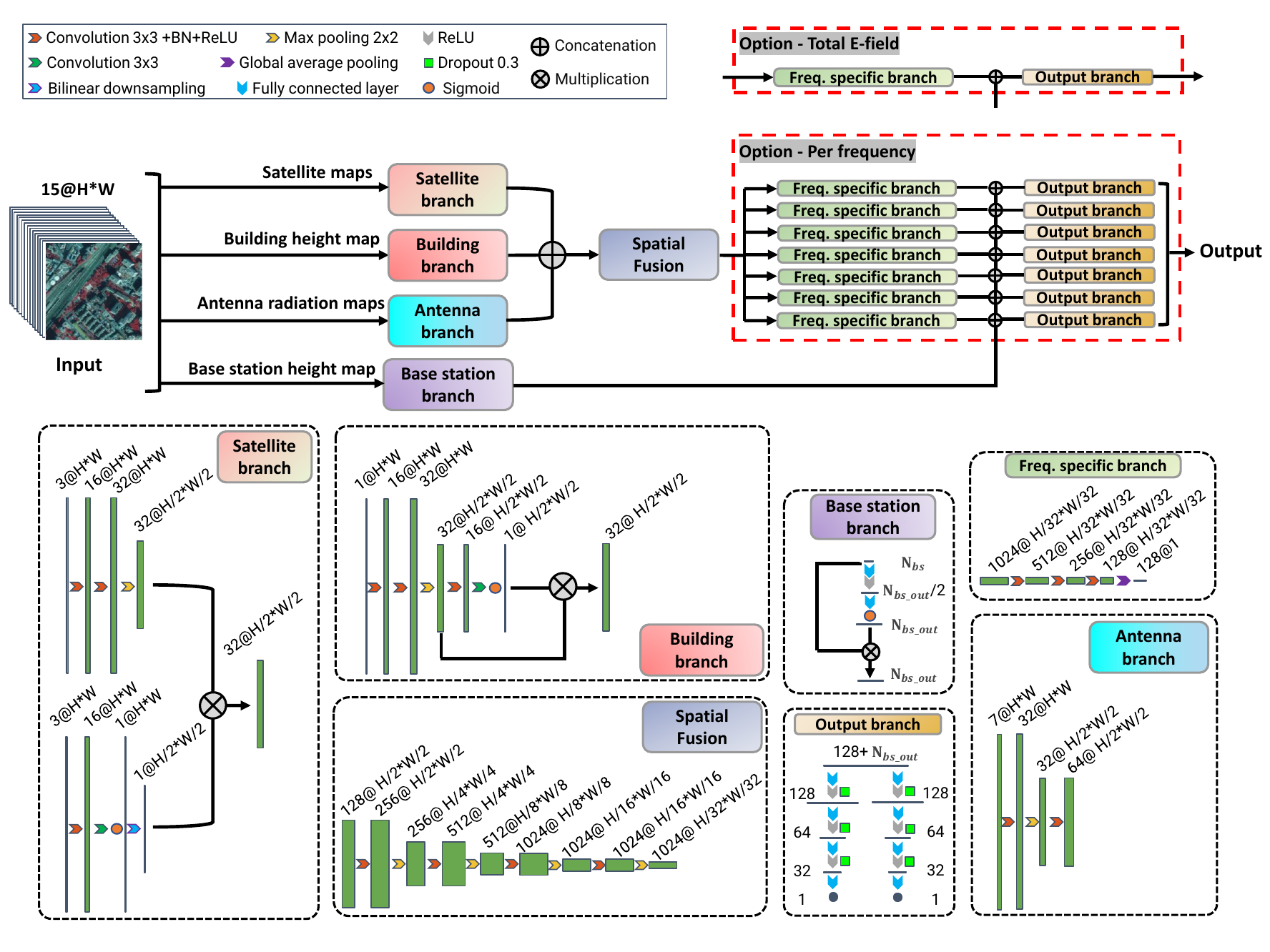}
    \caption{Overall architecture of ExposNet.}
    \label{fig:CNN}
\end{figure*}

\subsection{Network architecture}
The neural network must be carefully designed to handle multi-modal data as inputs, and to extract meaningful features to perform the regression task, specifically predicting the RMS and STD of the E-field within a given area. The network should not only be capable of extracting and combining features from different modalities but also preserve spatial information, so that features from different data modalities at the same location can be jointly analyzed.

Based on this concept, a deep neural network, named ExposNet, has been developed. The overall architecture is depicted in Fig.~\ref{fig:CNN}. Broadly, ExposNet consists of $7$ main modules: \textit{Satellite branch}, \textit{Building branch}, \textit{Antenna branch}, \textit{Base station branch}, \textit{Spatial fusion}, \textit{Frequency-specific branch}, and \textit{Output branch}. 
To make the structure easier to understand, we describe the network in three main phases:
\subsubsection{Input processing phase}
This phase focuses on pre-processing and extracting features from the multi-modal input data, ensuring that each modality is effectively represented for further integration.

In this phase, the \textit{Satellite branch} takes the infrared image and land cover map as input. The infrared image is first processed through two consecutive convolutional blocks, with the number of filters set to $16$ and $32$, respectively. Each convolutional block consists of a convolutional layer with a kernel size of $3$, a stride size of $1$, and one-pixel zero padding, followed by batch normalization (BN) and a ReLU activation layer. The configuration of the convolutional blocks remains consistent throughout the network. The resulting feature map is then downsampled using a max-pooling layer. In parallel, the land cover map is processed through a convolutional block, followed by a single convolutional layer and a sigmoid activation function, resulting in a single-channel feature map. This map is then downsampled using bilinear interpolation. The resulting feature map is used as a weight map that which is multiplied with the infrared image feature map. The objective of this step is to leverage the land cover map to enhance the feature extraction process by providing spatial guidance. The land cover map highlights different surface types, such as built-up areas, vegetation, and water, and the weight map emphasizes regions of importance for the infrared image features. By applying this weighting, the network can focus on surface-specific features, improving its ability to capture spatial and contextual information relevant to the regression task.

The output of the \textit{Satellite branch} is a $32$-channel weighted feature map derived from the satellite images. 

\textit{Building branch} takes the building height map as input. First, the input is processed through two consecutive convolutional blocks, followed by a max-pooling layer. The resulting feature map is then refined using a spatial attention mechanism. Specifically, the feature map is passed through a convolutional block, followed by a single convolutional layer and a sigmoid activation function, to produce a single-channel weighting map. This weighting map is then multiplied with the previous feature map, applying spatial weighting to emphasize regions that are more relevant to the task. This spatial attention mechanism allows the network to focus on areas with important spatial characteristics, such as buildings of significant height or clusters of buildings, which may have higher impact on EMF propagation. By doing so, the network ensures that key spatial information is retained and highlighted, while less relevant regions are de-emphasized.

The output of the \textit{Building branch} is also a $32$-channel weighted feature map, which captures and height-related information.

The \textit{Antenna branch} processes the antenna radiation maps across seven frequency bands as input. These inputs are passed through a convolutional block, followed by a max-pooling layer, and another convolutional block, resulting in a $64$-channel feature map.

The last module in this phase is the \textit{Base station branch}, which takes the BSA height map as input. Unlike other modules that focus on extracting spatial features, this branch emphasizes the height values themselves while disregarding spatial coordinates. This decision is based on several reasons: 
\begin{itemize}
    \item The BSA height map is inherently sparse, with only a few non-zero elements. Convolutional operations on such a sparse map are difficult to extract meaningful features.
    \item The location information of BSAs is already captured in the antenna radiation maps, so the height map primarily serves as a source of height-related values.
\end{itemize}

To achieve this, \textit{Base station branch} relies solely on fully connected layers to process the non-zero height values. First, the non-zero elements from the base station height map are extracted, and their total number is denoted as $N_{bs}$. These non-zero values are then fed into the branch as input. The processing steps involve passing the input through a fully connected layer, followed by a ReLU activation layer. The resulting features are passed through a second fully connected layer, which maps the features to a final dimension of $N_{bs\_out}$. Finally, a sigmoid activation function is then applied to normalize the output values, treating them as attention weights for the corresponding non-zero elements. The input height values are then expanded to match the dimensionality of the weights and are multiplied element-wise with the weights to compute weighted features. This approach ensures that even highly sparse inputs are effectively processed, as the network dynamically learns to focus on the most relevant non-zero values through attention-like weight computation. If there are no non-zero values in the input, i.e., an entirely empty base station height map for a given sample, a zero vector of size $N_{bs\_out}$ is returned for that sample.

\subsubsection{Spatial Feature Fusion Phase}
 After the preliminary feature extraction from the input channels, the outputs of \textit{Satellite branch}, \textit{Building branch} and \textit{Antenna branch} are concatenated and fed into the \textit{Spatial Fusion} module. This fusion module consists of four repeated combinations of convolutional blocks and max-pooling layers, resulting in a $1024$-channel feature map with a spatial resolution of $H/32 \times W/32$. The purpose of this module is not only to further extract meaningful features but also to effectively combine information across multiple modalities while preserving spatial relationships and connections. This ensures that the fused feature map captures both modality-specific characteristics and their interdependencies.
\subsubsection{Output Phase}

This phase is responsible for performing the final processing steps and generating prediction results. Notably, this phase offers two variants, which can either be used for frequency-selective predictions or total E-field predictions, depending on the nature of the measurement data. 

As illustrated in Fig.~\ref{fig:CNN}, the red dotted englobed parts highlight the output phase.
In ``Option - Per Frequency'', the output from the previous fusion phase is fed into seven \textit{Frequency specific branches}, each dedicated to one frequency band. Each branch consists of three convolutional blocks followed by a global average pooling layer, which reduces the spatial feature map to a feature vector with $128$ elements. This feature vector is then concatenated with the output from the \textit{Base station branch}, which is a vector of dimension $N_{bs\_out}$. The concatenated vector is then passed to the \textit{Output branch}, which consists of two parallel sub-branches. Each sub-branch is composed of fully connected layers with ReLU activation layers. A dropout rate of $0.3$ is consistently applied to enhance regularization and preventing overfitting. These sub-branches produce the final predictions: two scalar values representing the RMS and STD of the E-field level. This design, with seven parallel combinations of \textit{Frequency specific branches} and \textit{Output branches}, allows for frequency-selective predictions by independently processing and predicting results for each frequency band.

In ``Option - Total E-field'', only one combined branch is retained, which processes the fused feature map and base station features together to produce the RMS and STD predictions for the total E-field. This streamlined approach focuses on aggregate field predictions rather than individual frequency bands.
\section{Experimental Setup}\label{Section5}

\subsection{Dataset preparation}
As previously mentioned, the dataset includes $7516$ measurements from Paris and $19062$ from Lyon, amounting to a total of $26578$ measurements. In prior studies, the prediction target has typically been the exposure level at individual measurement points. However, since consecutive measurement points are only $2$ to $3$ meters apart, their recorded values are highly correlated and associated environmental features are nearly identical. To address this redundancy and enhance the generality of the predictions, we opted to predict the average E-field value within a defined area instead of focusing on point-wise predictions.

\begin{figure}[htbp]
    \centering
    \includegraphics[width=0.6\linewidth]{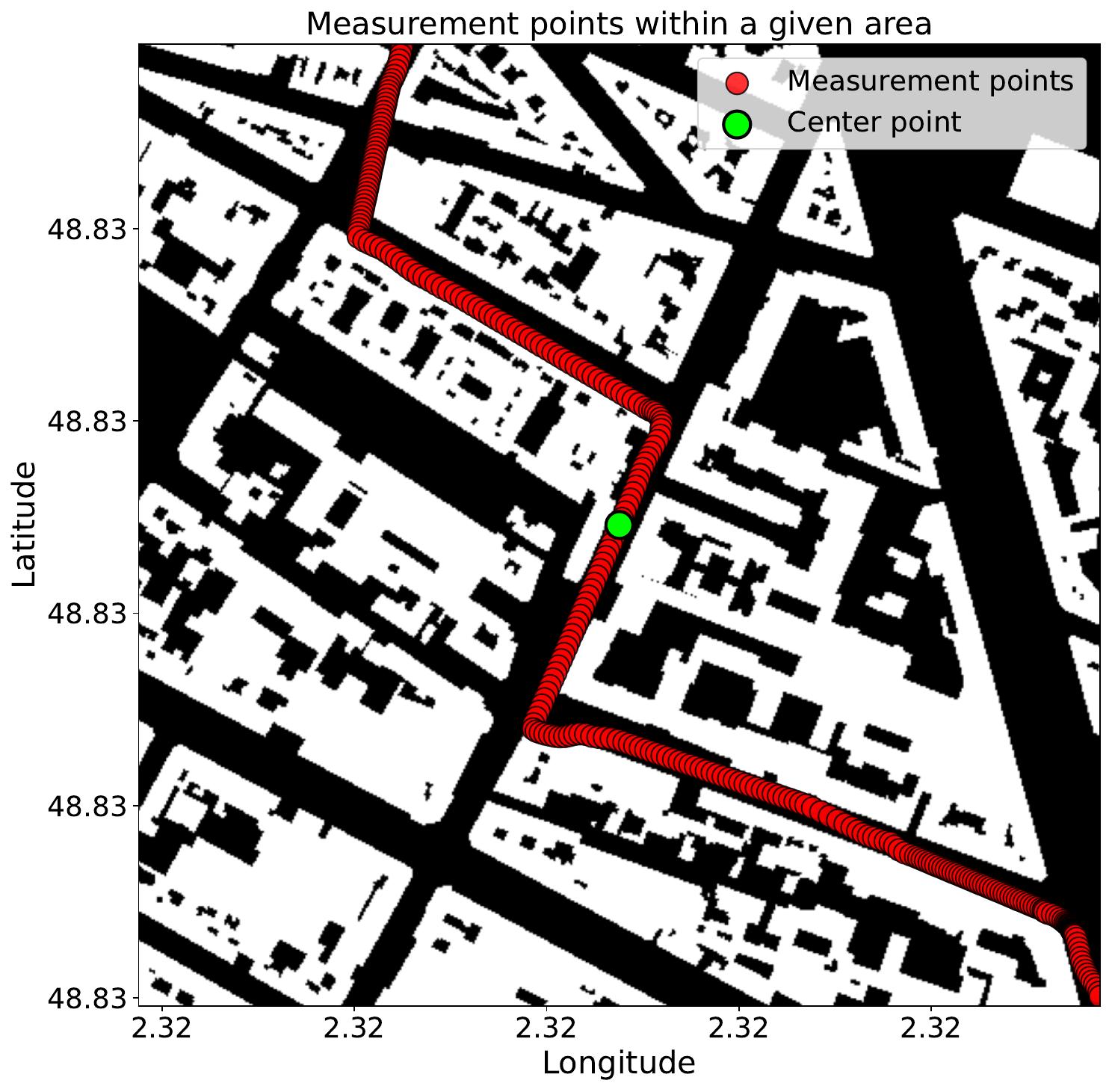}
    \caption{Example of a selected area.}
    \label{fig:PoinsInArea}
\end{figure}

The definition of the chosen areas is another a key consideration. The strategy adopted involves selecting areas with $50$-meter separation distance between adjacent area centers. All measurement points located within the corresponding area are then used to calculate the RMS and STD of E-field values, which constitute the model's outputs. The choice of the separation distance between area centers requires careful consideration. If the separation distance is too small, there will be excessive overlap between neighboring areas, leading to redundancy. Conversely, if the distance is too large, the number of areas, i.e., the dataset size, will be significantly reduced, potentially resulting in insufficient data for model training. Fig.~\ref{fig:PoinsInArea} illustrates an example of a selected area. The background represents the building height map, white points indicate all measurement locations within the area, and the green point marks the center point of the area.

By defining the area centers with a separation distance of $50$ meters, the obtained dataset contains $418$ samples for Paris and $1271$ for Lyon. Each area is defined as a square with the side length $N = $ \SI{400}{\metre}. The height and width of the image are $H = 128$ pixels and $W = 128$ pixels, respectively. In this setup, each pixel represents a real-world area of $3.125 \times 3.125$ \SI{}{\metre\squared}.

Regarding the division of the dataset to obtain training and testing subsets, several challenges need to be carefully addressed. First, given the small size of the dataset, we decided not to use a separate validation set. Additionally, since consecutive areas are very close to each other, the commonly used approach of randomly shuffling the dataset before splitting it into training and testing subsets is unsuitable. Such an approach would lead to significant overlap in information between the subsets, influencing the validity of the evaluation. To address this issue, we divided the dataset into two parts without shuffling. Specifically, $1539$ samples from Lyon and Paris are used for training, while the remaining $150$ consecutive samples from Paris are allocated for testing. This approach ensures that the training and testing datasets are distinct.
\subsection{Loss function and evaluation metric}
The Mean Square Error (MSE) loss is employed during the training stage. In the following, the loss function and evaluation metrics used for ``Option - Per frequency'' and ``Option - Total E-field'' are introduced, respectively. 

\subsubsection{``Option - Per frequency''}
For frequency-selective prediction, the loss function consists of two components: the data fidelity term and the constraint term. The data fidelity term is expressed as:
\begin{equation}
    \mathcal{D} = \dfrac{1}{N_{s}\times N_f \times 2}\sum^{N_{s}}_{i=1}\sum^{N_f}_{j=1} \sum^2_{k=1} (y_{\text{pred}}[i,j,k]-
    y_{\text{true}}[i,j,k])^2
\end{equation}
where $i \in [1, N_{s}]$ is the sample index, $j \in [1,N_f]$ (with $N_f = 7$) is the frequency band index, and $k=1,2$ corresponds to RMS or STD value, respectively.

The constraint term is defined as:
\begin{equation}
 \mathcal{C} = \dfrac{1}{N_s}\sum^{N_s}_{i=1}\left(E_{\text{total\_pred}}[i]-E_{\text{total\_true}}[i]\right)^2
\end{equation}
where the predicted and true total E-field values are calculated as follows:
\begin{align}
E_{\text{total\_pred},i} &= \sqrt{\sum^{N_f}_{j=1}(y_{\text{pred}}[i,j,1])^2}\\
E_{\text{total\_true},i} &= \sqrt{\sum^{N_f}_{j=1}(y_{\text{true}}[i,j,1])^2}
\end{align}
The final loss function combines these two terms as:
\begin{equation}
    \mathcal{L}_1 = \mathcal{D}+\lambda\cdot \mathcal{C}
\end{equation}
where $\lambda = 0.1$ is a hyperparameter that balances the trade-off between the data fidelity term and the constraint term. 

The purpose of the constraint term is to ensure consistency between the predicted frequency-specific E-field values and the total E-field. Without this term, the model may focus solely on minimizing the errors for individual frequency bands, potentially causing the total E-field derived from these predictions to deviate from the true total E-field. This constraint term forces the model to not only optimize the per-band predictions but also to learn the overall relationship among the $7$ frequency bands, maintaining global consistency while reducing band-specific errors. Here, the constraint is only based on the discrepancy between the predicted and true RMS of the total E-field values, which are computed from the predicted and true RMS of per-band E-field values using Eq.~\eqref{EtotalCalculation}. It does not involve the STD, as the STD of the total E-field values cannot be derived from the RMS of per-band E-field values without knowing the full distribution of the E-field across all measurement points within the given area. 

The evaluation metrics used to assess model performance are the Root Mean Squared Error (RMSE) and Mean Absolute Percentage Error (MAPE), which are defined as:
\begin{equation}
    \text{RMSE}_{j, k} = \sqrt{\dfrac{1}{N_s} \sum^{N_s}_{i=1} (y_{\text{pred}}[i,j,k] - y_{\text{true}}[i,j,k])^2}
\end{equation}
and
\begin{equation}
    \text{MAPE}_{j, k} = \dfrac{1}{N_s} \sum^{N_s}_{i=1} \left|\dfrac{y_{\text{pred}}[i,j,k] - y_{\text{true}}[i,j,k]}{y_{\text{true}}[i,j,k]}\right| \times 100 \%
\end{equation}
\subsubsection{``Option - Total E-field''}
For total E-field prediction, the loss function simplifies to the MSE loss between the predicted and true total E-field values:
\begin{equation}
    \mathcal{L}_2 =  \dfrac{1}{N_{s} \times 2}\sum^{N_{\text{s}}}_{i=1} \sum^2_{k=1} (y_{\text{pred}}[i,k]- y_{\text{true}}[i,k])^2
\end{equation}
In this case, the evaluation metrics $\text{RMSE}_{j,k}$ and $\text{MAPE}_{j,k}$ are reduced to $\text{RMSE}_{k}$ and $\text{MAPE}_{k}$, as the frequency band index $j$ is no longer applicable.

\subsection{Implementation details}
The implementation details of the training process are as follows:
\begin{itemize}
    \item Input normalization: The input data are normalized using the min-max normalization method to scale the values between $0$ and $1$.
    \item Optimizer and weight decay: The Adam optimizer is used for training, with different weight decay values depending on the prediction type. For frequency-selective prediction, the weight decay is set to $1\times 10^{-4}$, while for total E-field prediction, it is set to $1\times 10^{-5}$.
    \item Training configuration: The training is conducted over $40$ epochs, with a mini-batch size of $8$. The initial learning rate is set to $1\times 10^{-4}$ and is reduced by a factor of $0.5$ every $5$ epochs.
    \item Training environment: The network is implemented using PyTorch and trained on a PC equipped with an NVIDIA T$600$ Laptop GPU (\SI{4}{\giga\byte}).
    \item Training time: For frequency-selective prediction, training takes approximately $24.6$ minutes, while for total E-field prediction, it takes $18.6$ minutes.
\end{itemize}

\section{Results and discussions}\label{Section6}
\begin{figure*}[htbp]
    \centering
    \includegraphics[width=0.45\linewidth]{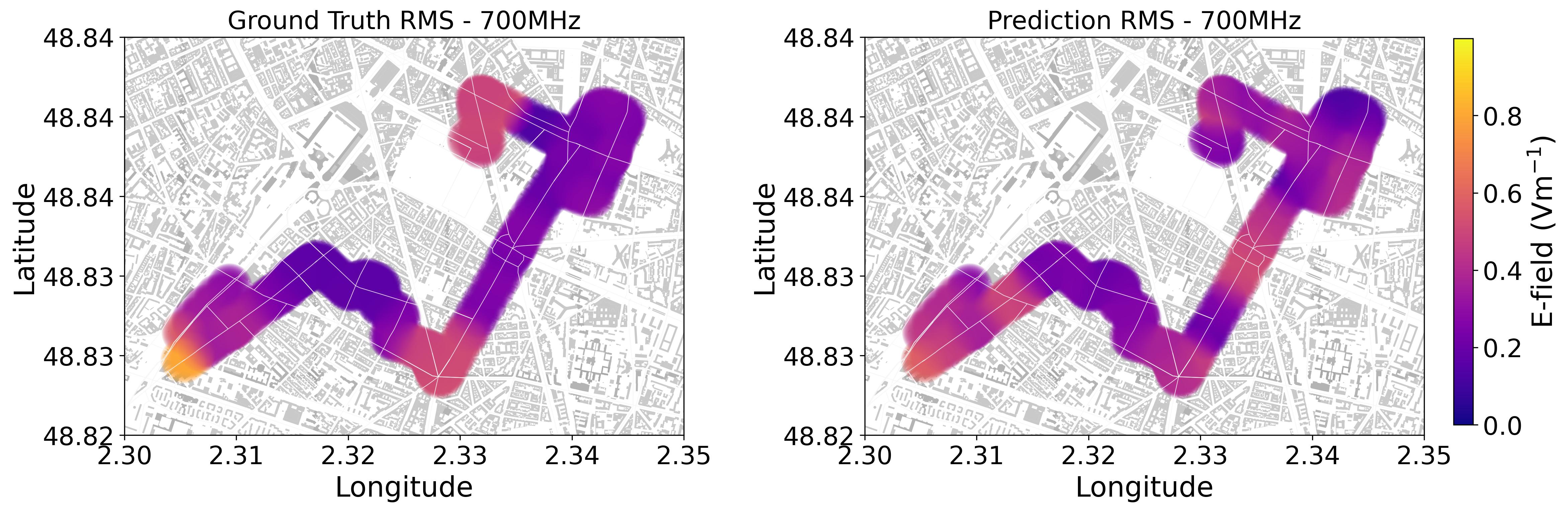}
    \includegraphics[width=0.45\linewidth]{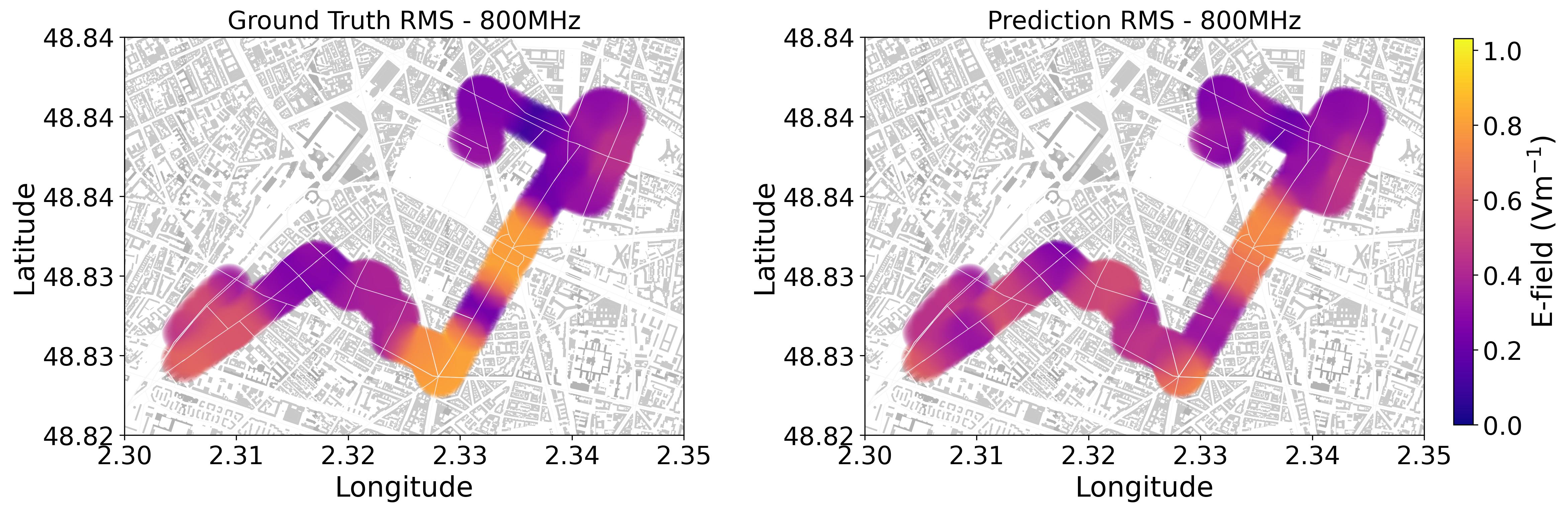}
    
    \includegraphics[width=0.45\linewidth]{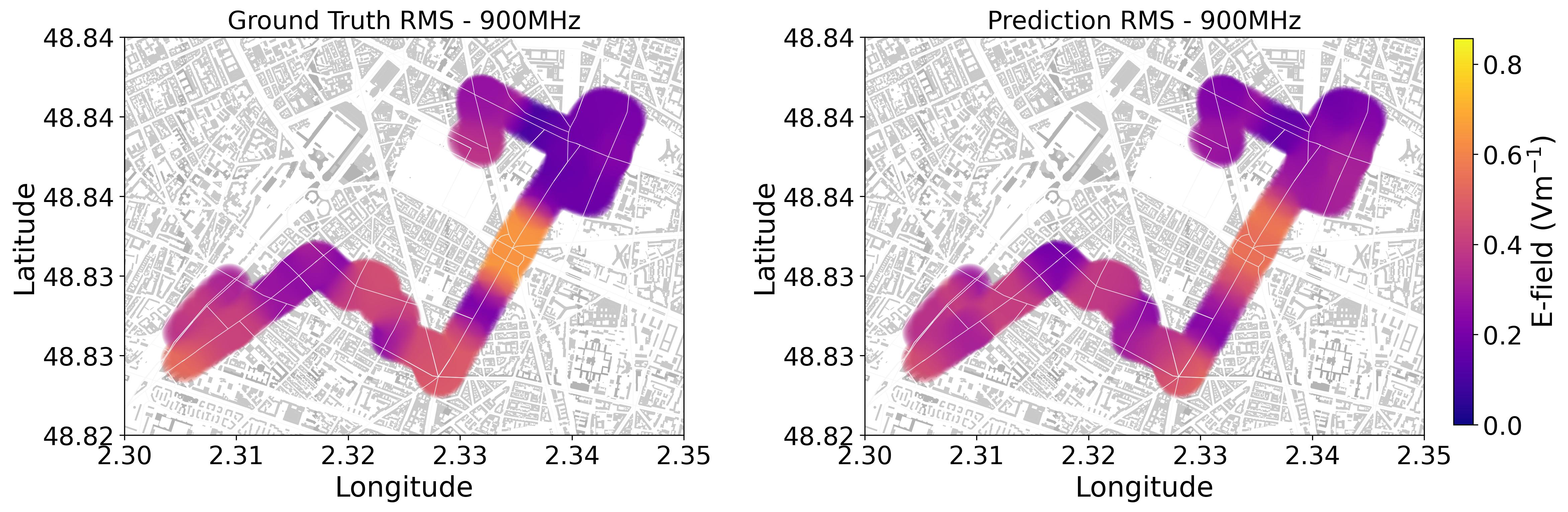}
    \includegraphics[width=0.45\linewidth]{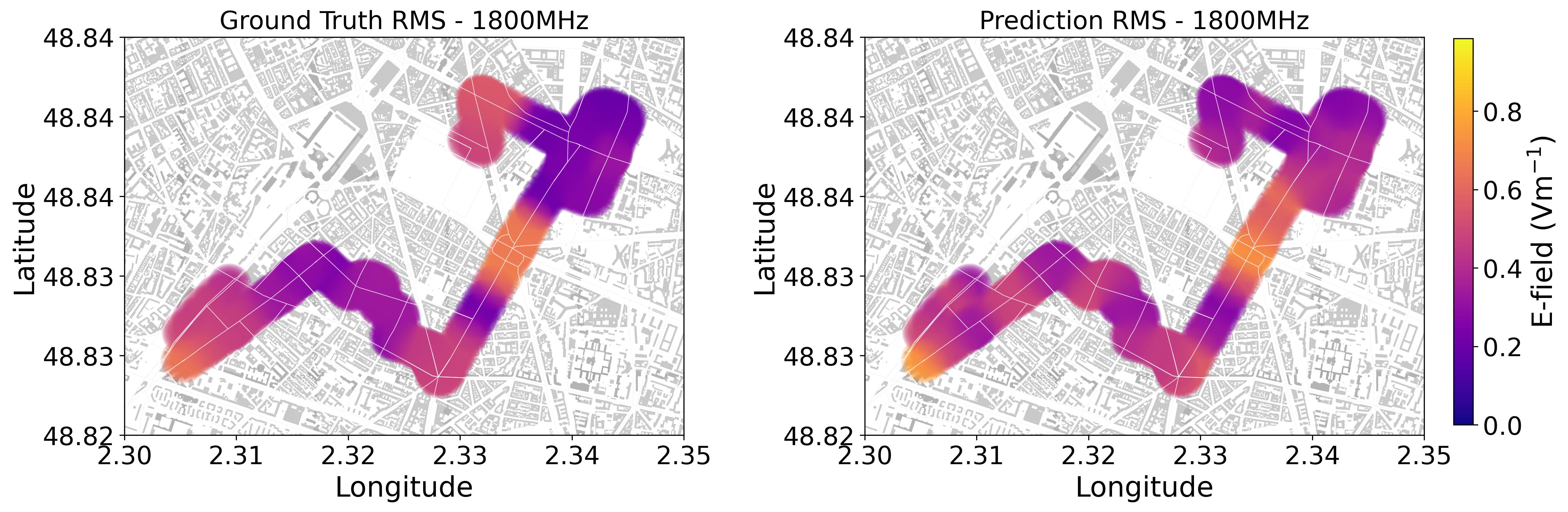}

    \includegraphics[width=0.45\linewidth]{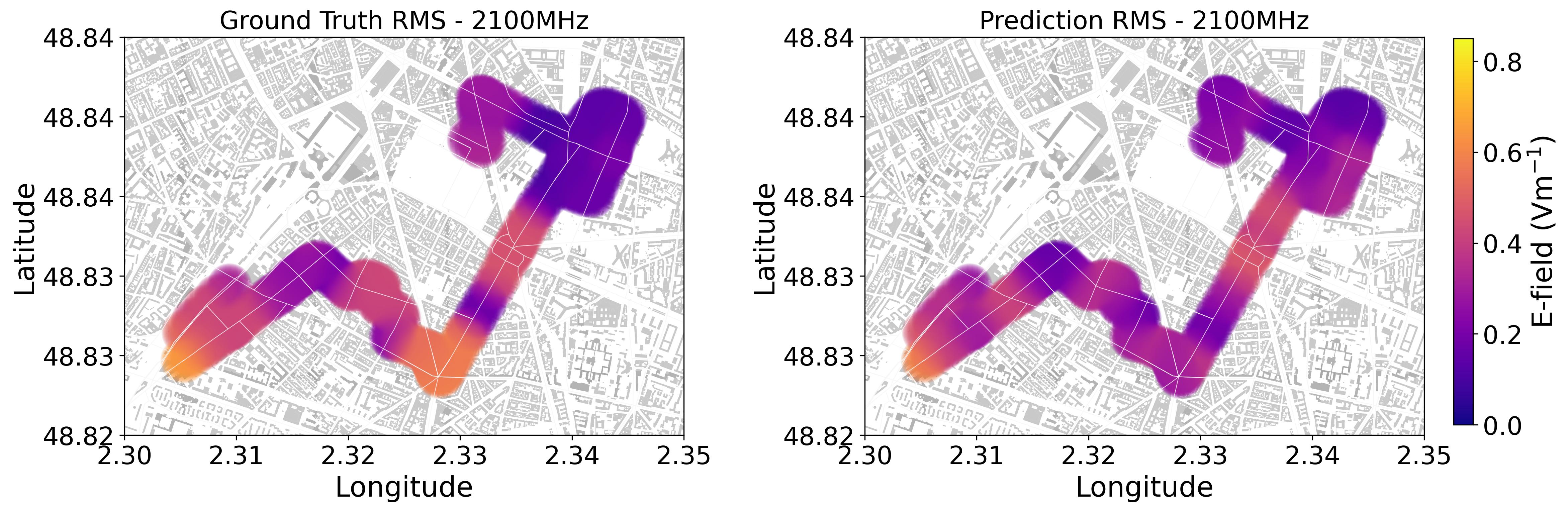}
    \includegraphics[width=0.45\linewidth]{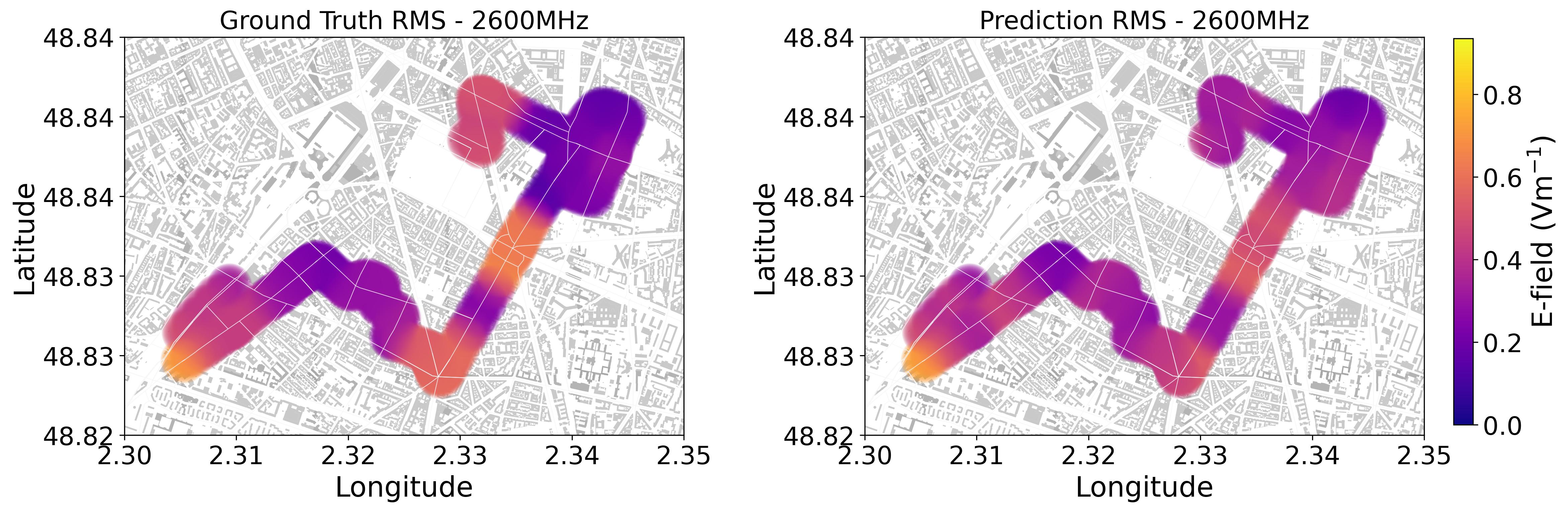}
    
    \includegraphics[width=0.45\linewidth]{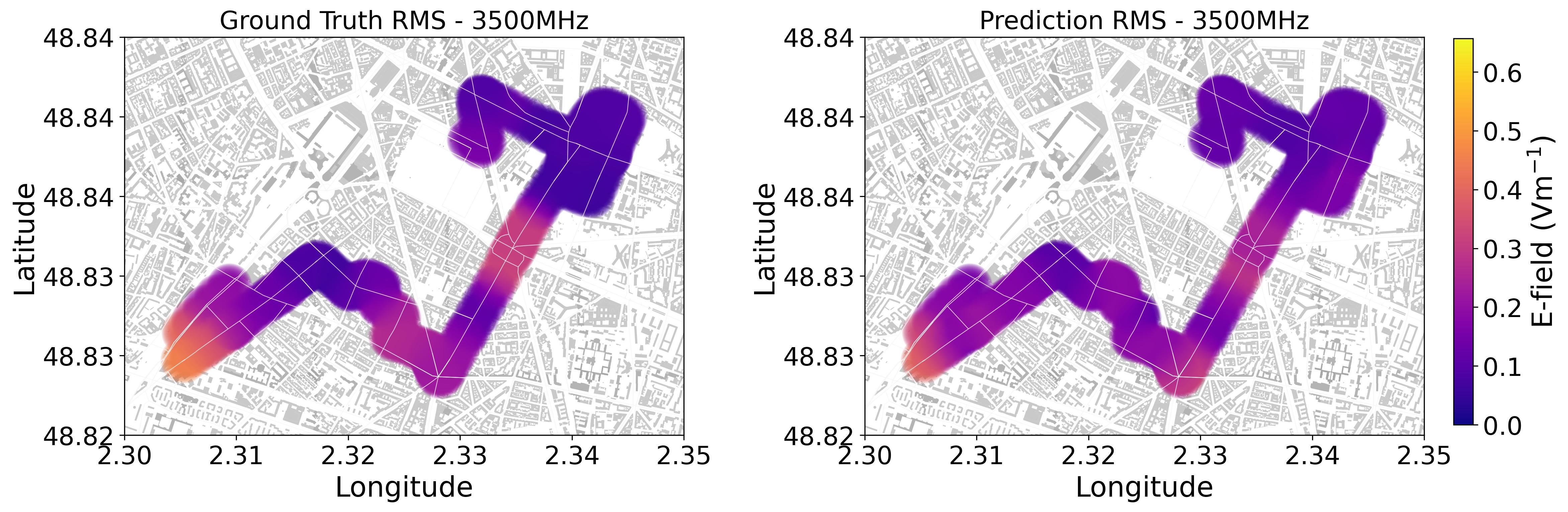}
    \includegraphics[width=0.45\linewidth]{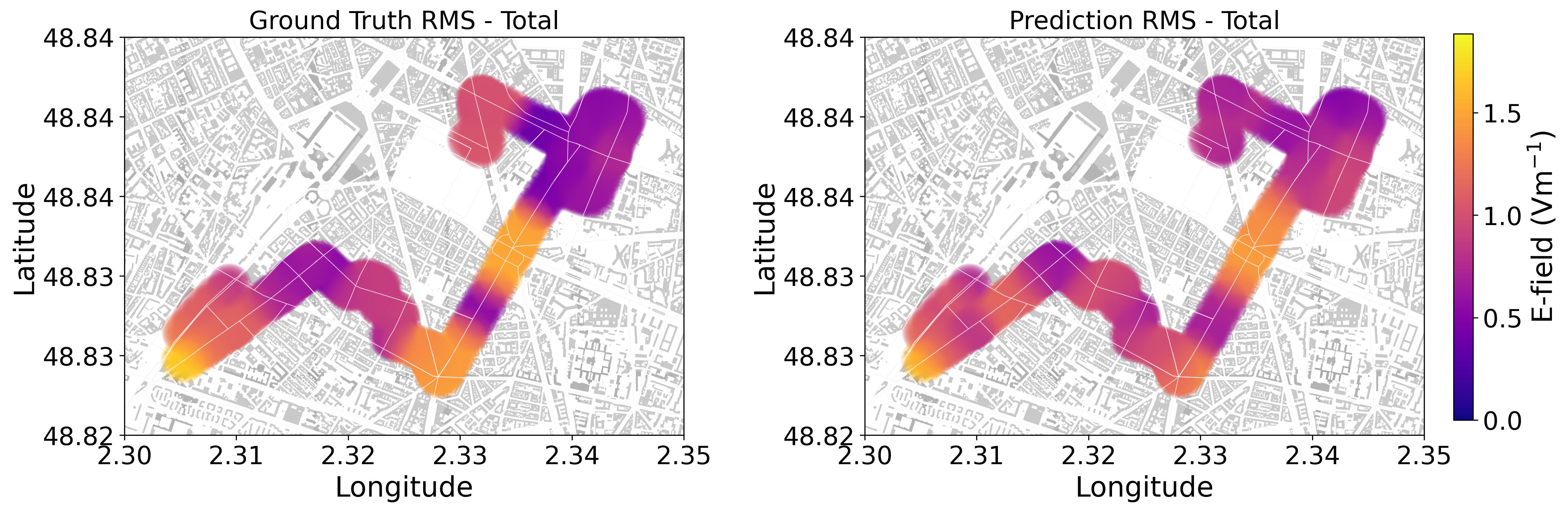}  
    
    \caption{Results visualization and comparison for ``Option - Per Frequency''.}
    \label{fig:MappingResults1}
\end{figure*}

In this section, the performance of the model will be analyzed and discussed respectively for frequency-selective prediction and total E-field prediction.
\subsection{Frequency-selective prediction}
First, we choose ``Option - Per Frequency'' to perform frequency-selective predictions. The trained model is then evaluated using $150$ test samples obtained from measurements in Paris. Table~\ref{tab:1} presents the RMSE values for the predicted RMS and STD across all frequency bands, along with the RMS of the total E-field, which is calculated from the per-frequency predictions using Eq.~\eqref{EtotalCalculation}. Since the STD of the total E-field cannot be directly obtained from the per-frequency STD values, only the RMS of the total E-field is reported.
\begin{table}[hbtp]
\centering
\caption{RMSE (\SI{}{\volt\per\meter}) of per-frequency predictions:~Testing dataset from Paris }
\label{tab:1}
\resizebox{\columnwidth}{!}{%
\begin{tabular}{|c|c|c|c|c|c|c|c|c|}
\hline
Freq. (\SI{}{\mega\hertz})& 700   & 800   & 900   & 1800  & 2100  & 2600  & 3500  & Total \\ \hline
RMS         & 0.129 & 0.149 & 0.085 & 0.125 & 0.130 & 0.102 & 0.058 & 0.241 \\ \hline
STD         & 0.077 & 0.099 & 0.052 & 0.071 & 0.083 & 0.078 & 0.046 & N/A   \\ \hline
\end{tabular}%
}
\end{table}

From Table \ref{tab:1}, it can be observed that the RMSE for RMS prediction across all frequency bands remains well below \SI{0.15}{\volt\per\meter}, with the lowest value of \SI{0.058}{\volt\per\meter} found in the \SI{3500}{\mega\hertz} band. Regarding the RMSE of STD prediction, values across all frequency bands range from $0.046$ to \SI{0.099}{\volt\per\meter}. For the computed total E-field RMS, the RMSE is \SI{0.241}{\volt\per\meter}. The RMSE for total E-field prediction is noticeably higher compared to per-frequency results, as the total field inherently has larger magnitudes than individual frequency components.

\begin{figure}[H]
    \centering
    \includegraphics[width=0.6\linewidth]{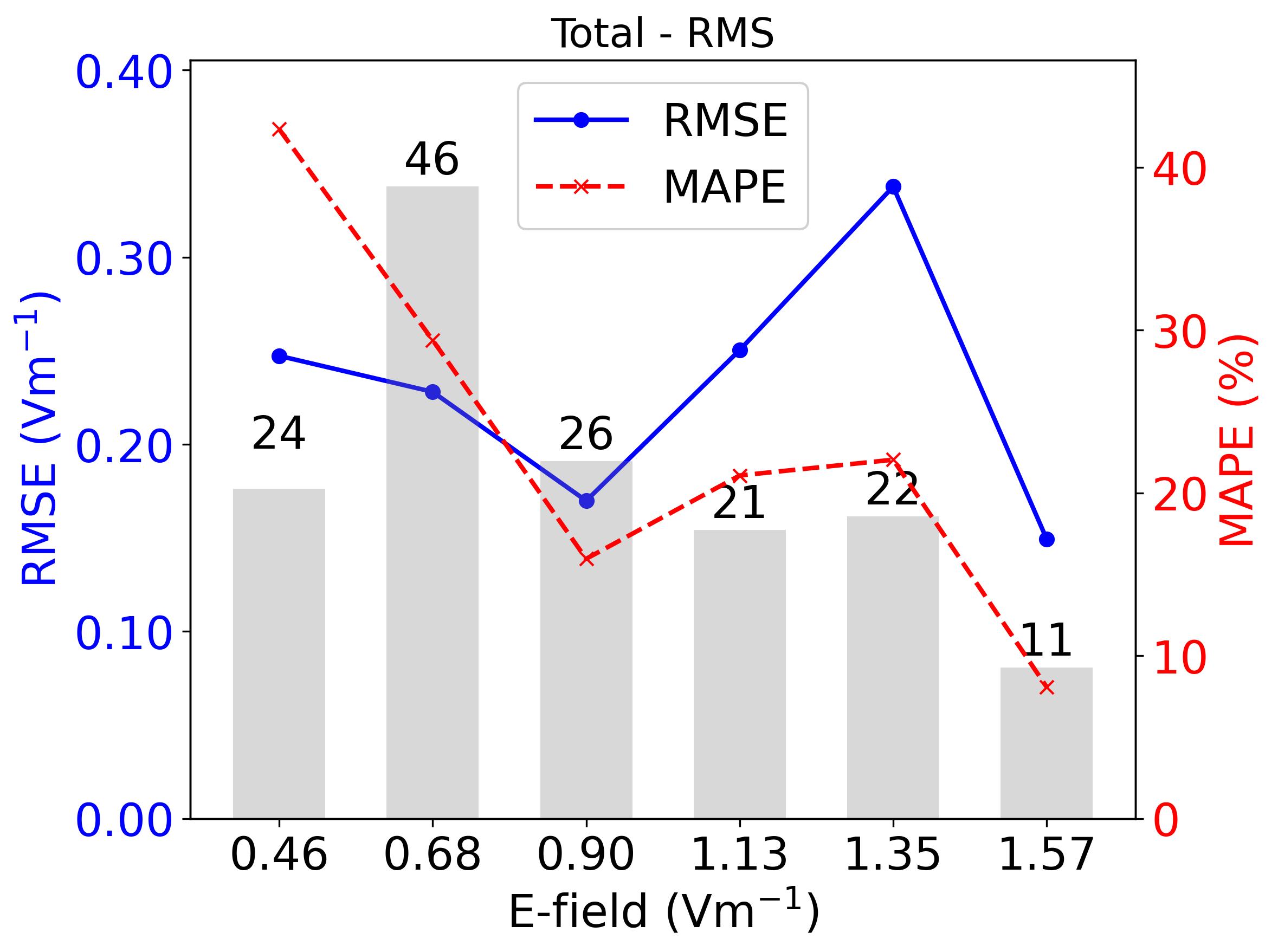}  
    \caption{MSE and MAPE across different magnitudes of total E-field values (``Option - Per Frequency'').}
    \label{fig:Error1}
\end{figure}

A more detailed analysis of total E-field prediction errors is provided in Fig.~\ref{fig:Error1}. 

In Fig. \ref{fig:Error1}, the x-axis represents the true E-field level. To better analyze RMSE and MAPE across different E-field amplitudes, the total dataset is divided into $6$ equally spaced intervals. Each interval is defined by partitioning the range between the maximum and minimum E-field values into $6$ equal segments. The values shown on the x-axis below each bar represent the median in each interval, rather than the maximum or minimum values across all samples. This ensures that the plotted RMSE and MAPE values provide a more representative measure of the central tendency within each interval.

The grey bars and the labels on top indicate the number of samples within each interval. The blue and red curves represent the RMSE and MAPE trends across the intervals, with the left y-axis corresponding to RMSE and the right y-axis corresponding to MAPE. From the figure, it can be observed that MAPE exhibits a generally decreasing trend. This is expected, as smaller E-field values tend to produce higher MAPE values, given that the loss function used for training is based on MSE. The overall MAPE average is $25.32\%$.  Regarding RMSE, the interval centered at 1.35 V/m exhibits the highest RMSE, exceeding \SI{0.30}{\volt\per\meter}, while in all other intervals, RMSE remains well below \SI{0.30}{\volt\per\meter}.

Fig.~\ref{fig:MappingResults1} presents a visualization of the ground truth and predicted values obtained using the ``Option - Per Frequency'' setting. Since the model predicts the average E-field level within a given area, the mapping is not represented as individual points but rather as a uniform value assigned to each area of $400 \times 400$ \SI{}{\metre\squared}. The results demonstrate that the model effectively captures the overall trends and spatial distribution of E-field levels across different magnitudes, both for individual frequency bands and for the total E-field level.

\subsection{Total E-field prediction}
In this subsection, we use the ``Option - Total E-field'' setting to generate predictions exclusively for the total E-field level. This approach is particularly suitable for datasets where frequency-selective results are not available. 

\begin{table}[htbp]
\centering
\caption{RMSE and MAPE of total E-field predictions:~Testing dataset from Paris}
\label{tab:2}
\begin{tabular}{|c|c|c|}
\hline
Target\textbackslash{}Metric & RMSE  & MAPE (\%) \\ \hline
RMS                          & 0.231 & 20.353     \\ \hline
STD                          & 0.153 & 27.127     \\ \hline
\end{tabular}%
\end{table}

The RMSE and MAPE evaluation results are presented in Table~\ref{tab:2}. Compared to the total E-field level derived from frequency-selective predictions, the directly predicted total E-field demonstrates lower overall RMSE and MAPE values for RMS prediction, indicating improved accuracy.
\begin{figure}[hbtp]
    \centering
    \includegraphics[width=0.98\linewidth]{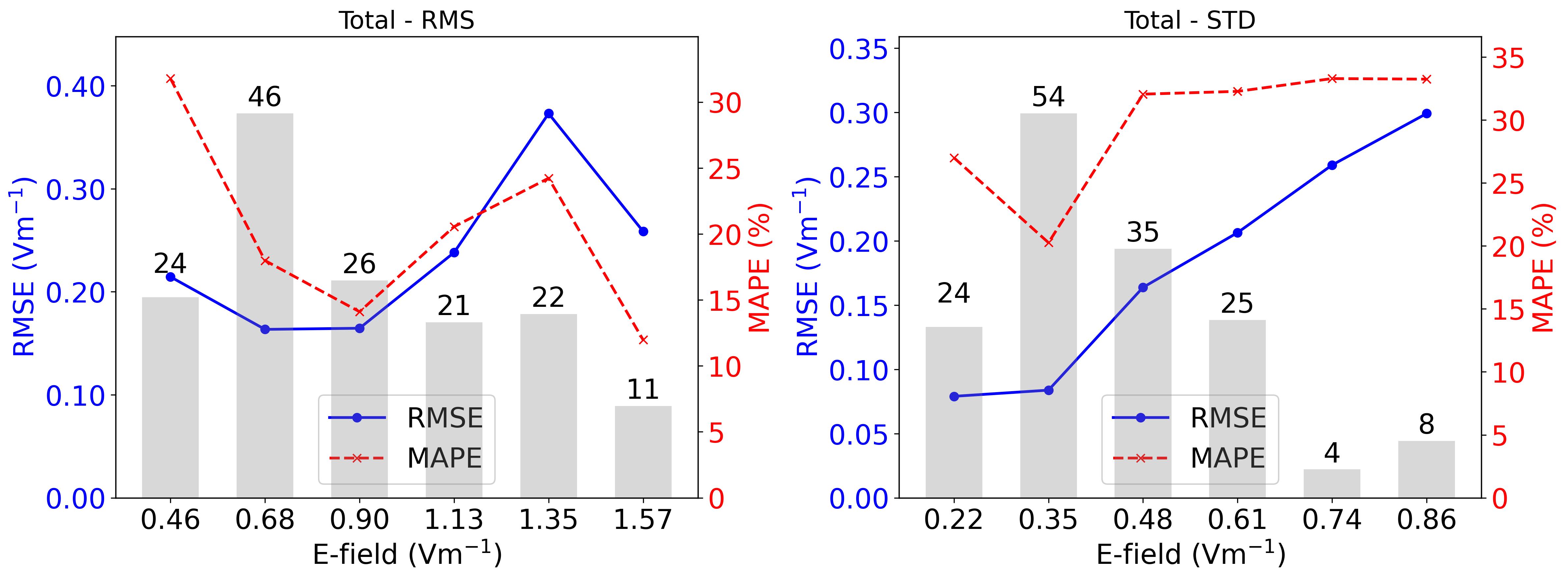}
    \caption{MSE and MAPE across different magnitudes of total E-field values (``Option - Total E-field'').}
    \label{fig:Error2}
\end{figure}

Fig.~\ref{fig:Error2} illustrates the variation of RMSE and MAPE across different E-field intervals for both RMS and STD predictions. The trends observed are similar to those presented in the frequency-selective results. Specifically, for lower E-field values, RMSE values remain relatively small, whereas for higher E-field intervals, MAPE values decrease. Regarding STD prediction, due to the generally low magnitudes, the MAPE of the predictions is consistently higher but remains below $28\%$ on average. Similarly, RMSE increases as E-field levels rise, which is reasonable.
\begin{figure}[hbtp]
    \centering
    \includegraphics[width=1\linewidth]{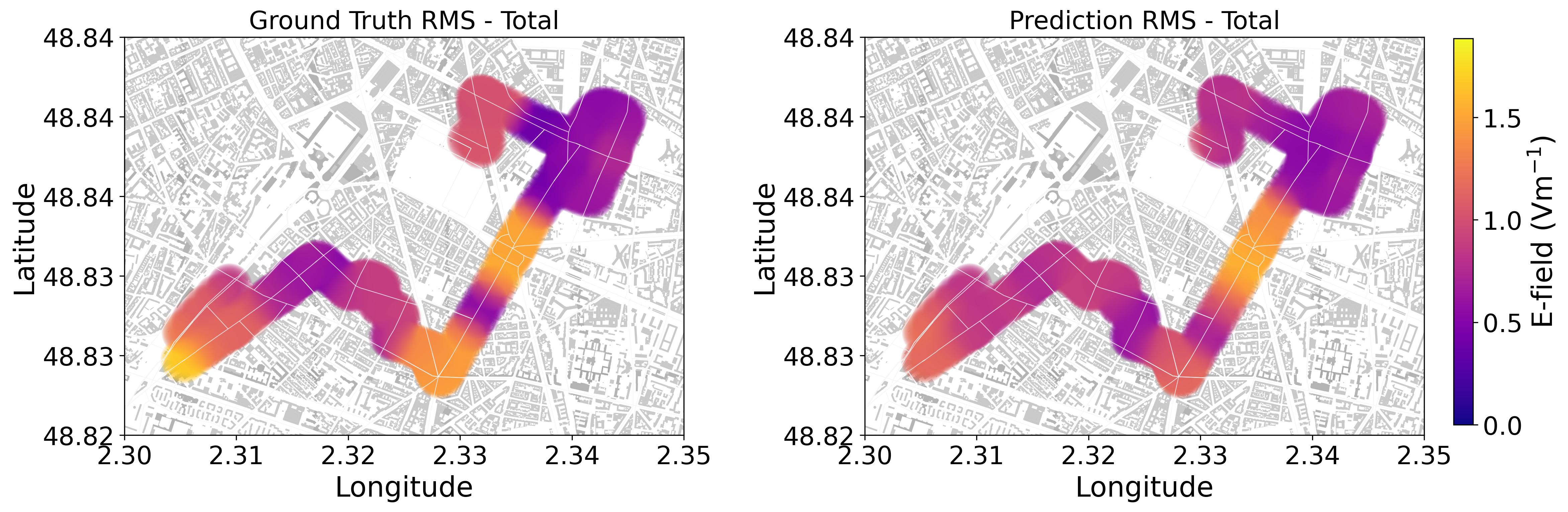}
    \caption{Results visualization and comparison for ``Option - Total E-field''.}
    \label{fig:MappingResults2}
\end{figure}

Fig.~\ref{fig:MappingResults2} provides a visualization of the prediction results. It can be observed that the model effectively captures and reconstructs the spatial distribution of E-field levels while successfully differentiating between areas with higher and lower E-field intensities.

\subsection{Generalizability evaluation}
It should be emphasized that a model's generalization ability is crucial for ensuring its robustness across different environments. In this experiment, rather than testing the trained model on measurements from Paris, we evaluate its performance using two separate testing datasets derived from measurements in Lyon, each consisting of $150$ test samples. The model is trained separately for each evaluation using the corresponding remaining dataset. These two datasets represent distinct scenarios, which will be analyzed in detail in the following parts.

\subsubsection{Testing dataset 1 from Lyon} 

\begin{table}[hbtp]
\centering
\caption{RMSE (\SI{}{\volt\per\meter}) of per-frequency predictions:~Testing dataset $1$ from Lyon}
\label{tab:3}
\resizebox{\columnwidth}{!}{%
\begin{tabular}{|c|c|c|c|c|c|c|c|c|}
\hline
Freq. (\SI{}{\mega\hertz})& 700   & 800   & 900   & 1800  & 2100  & 2600  & 3500  & Total \\ \hline
RMS         & 0.223 & 0.211 & 0.148 & 0.172 & 0.109 & 0.131 & 0.112 &  0.339\\ \hline
STD         & 0.151 & 0.162 & 0.098 & 0.113 & 0.066 & 0.096& 0.057& N/A   \\ \hline
\end{tabular}%
}
\end{table}

The first dataset from Lyon contains some relatively high E-field values, which are not frequent in the dataset. Table~\ref{tab:3} presents the RMSE values for per-frequency prediction results. It can be observed that the highest RMSE for RMS prediction occurs at \SI{700}{\mega\hertz}, reaching \SI{0.223}{\volt\per\meter}. The RMSE of the total E-field, estimated from per-frequency predictions, is approximately \SI{0.339}{\volt\per\meter}.  

\begin{figure}[H]
    \centering
    \includegraphics[width=0.56\linewidth]{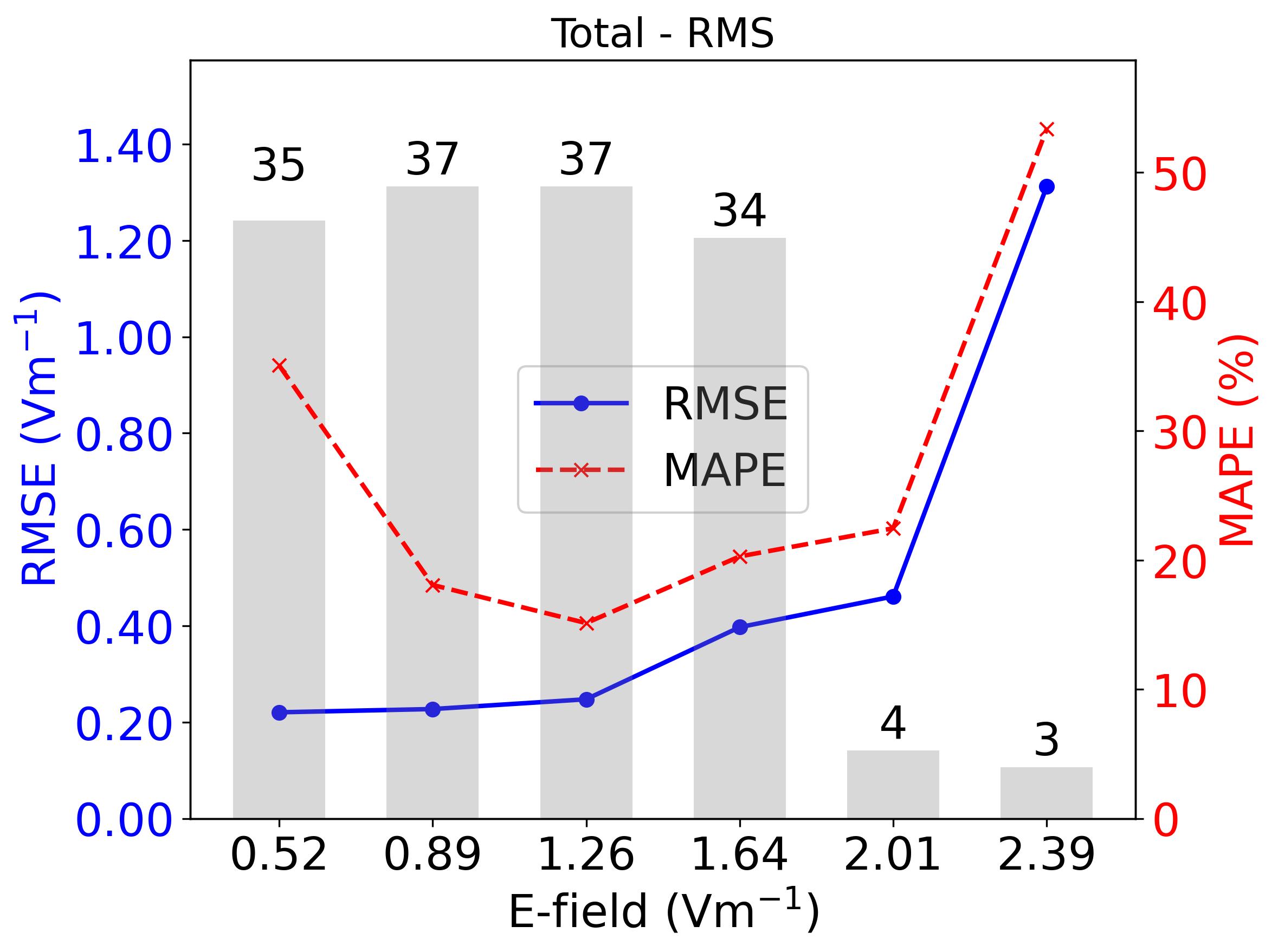}
    \includegraphics[width=0.42\linewidth]{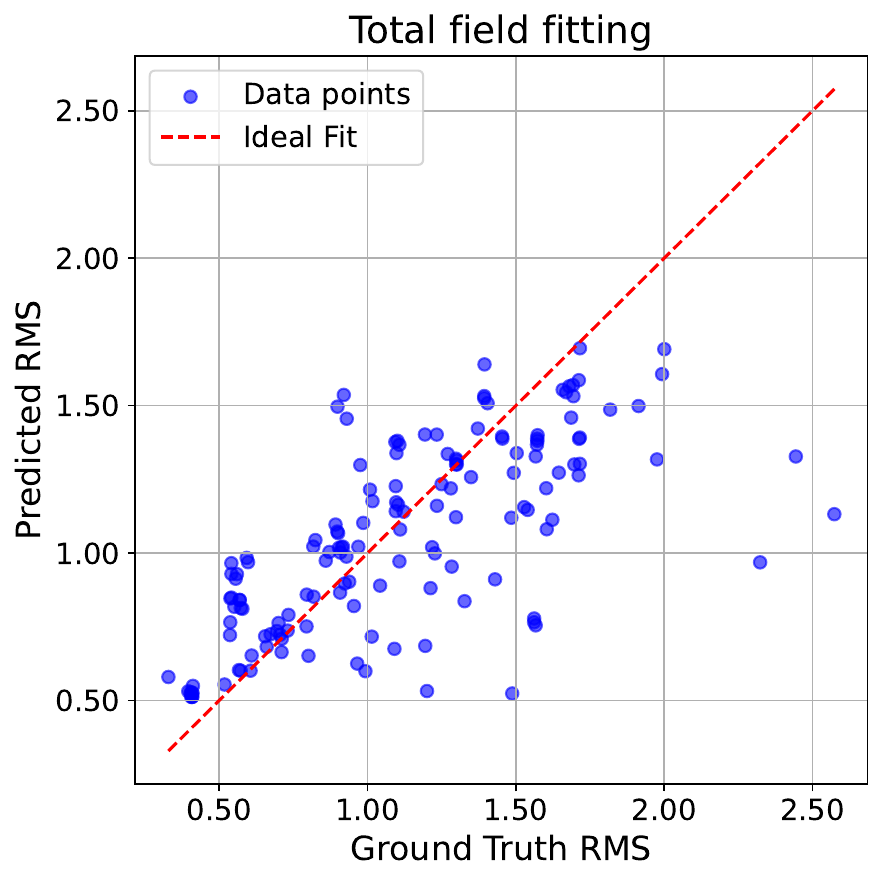}
    \caption{Results of testing dataset $1$ from Lyon: Left: MSE and MAPE across different magnitudes of total E-field values; Right: Scatter plot of predicted RMS vs. ground truth RMS of the total E-Field. }
    \label{fig:Error3}
\end{figure}

\begin{figure}[hbtp]
    \centering
    \includegraphics[width=1\linewidth]{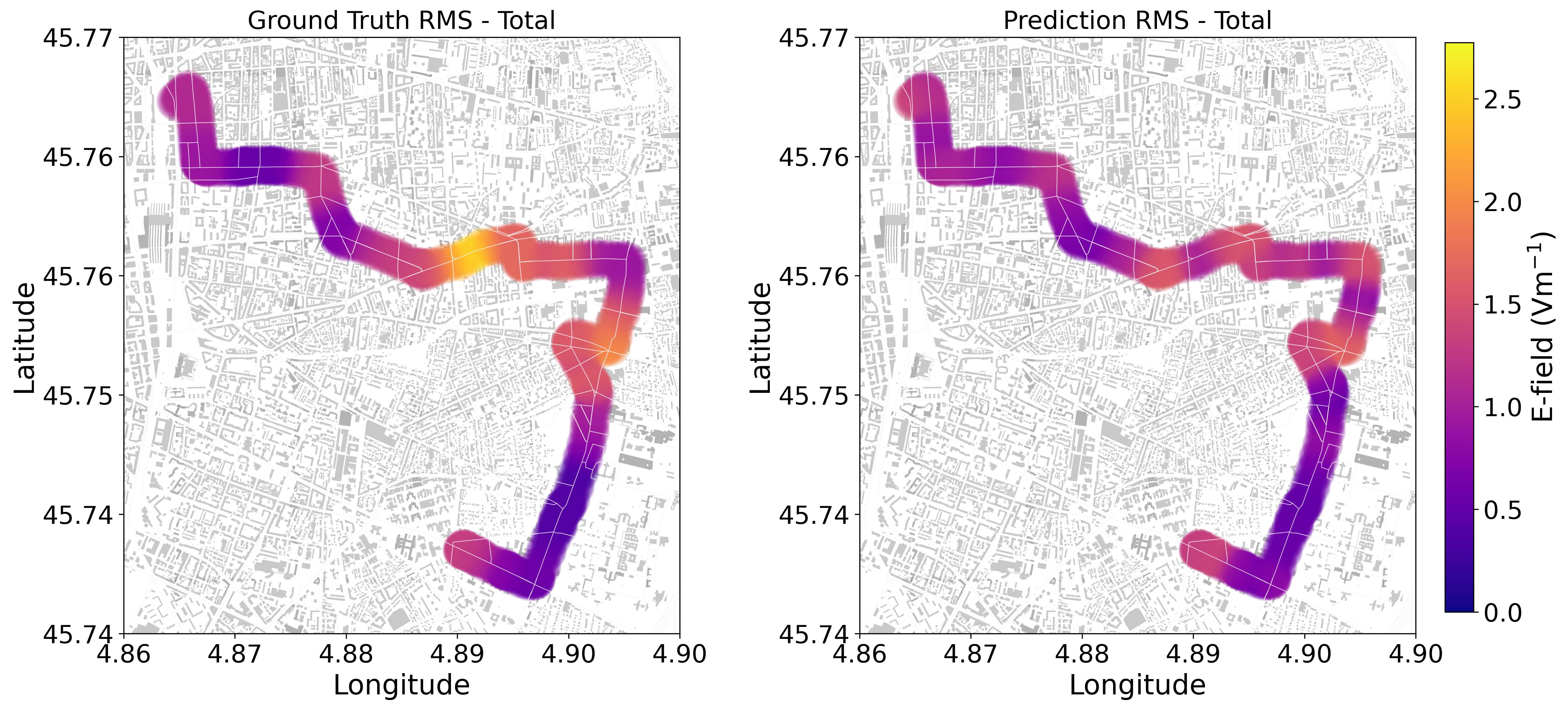}
    \caption{Results visualization and comparison for testing dataset $1$ from Lyon.}
    \label{fig:MappingResults3}
\end{figure}

From both the interval distribution in the bar plot and the scatter plot shown in Fig.~\ref{fig:Error3}, it can be seen that this dataset represents a segment of the route with overall higher E-field values. Furthermore, not only are the E-field values generally high, but the range of distribution is also broader. The scatter plot and the mapping results in Fig.~\ref{fig:MappingResults3} further confirm this observation. Notably, the areas with the highest E-field values are not well recognized by the model, affecting approximately 4 to 5 samples, as indicated in the scatter plot.  

The lower accuracy in predicting high values is reasonable, as the majority of samples in the training dataset are concentrated within a relatively narrow E-field range. As a result, the trained model struggles to accurately identify ``outlier'', i.e., areas with higher E-field values. These high E-field values are not frequent in the training dataset, making it difficult for the trained model to generalize effectively. An effective way to improve this issue is to increase the diversity of training samples, ensuring that higher E-field regions are better represented in the dataset. This could be achieved by incorporating additional measurement points from areas with strong E-field variations.

Nonetheless, apart from the few high E-field areas, the scatter plot and mapping results in Fig.~\ref{fig:Error3} indicate that the model successfully reconstructs the E-field levels for the majority of the samples, demonstrating overall good predictive performance. Additionally, adopting an adaptive loss function, such as a weighted loss that assigns higher importance to samples with extreme values, could help the model learn these cases more effectively. 

\subsubsection{Testing dataset 2 from Lyon}
The second dataset from Lyon represents a segment of the route situated in the city center, with relatively narrow streets and densely packed commercial zones. Here, we use ``Option - Total E-Field'' to directly predict the total E-field value.  
\begin{table}[htbp]
\centering
\caption{RMSE and MAPE of total E-field predictions:~Testing dataset $2$ from Lyon}
\label{tab:4}
\begin{tabular}{|c|c|c|}
\hline
Target\textbackslash{}Metric & RMSE  & MAPE (\%) \\ \hline
RMS                          & 0.265 & 21.200     \\ \hline
STD                          & 0.143 & 23.730    \\ \hline
\end{tabular}%
\end{table}

\begin{figure}[hbtp]
    \centering
    \includegraphics[width=1\linewidth]{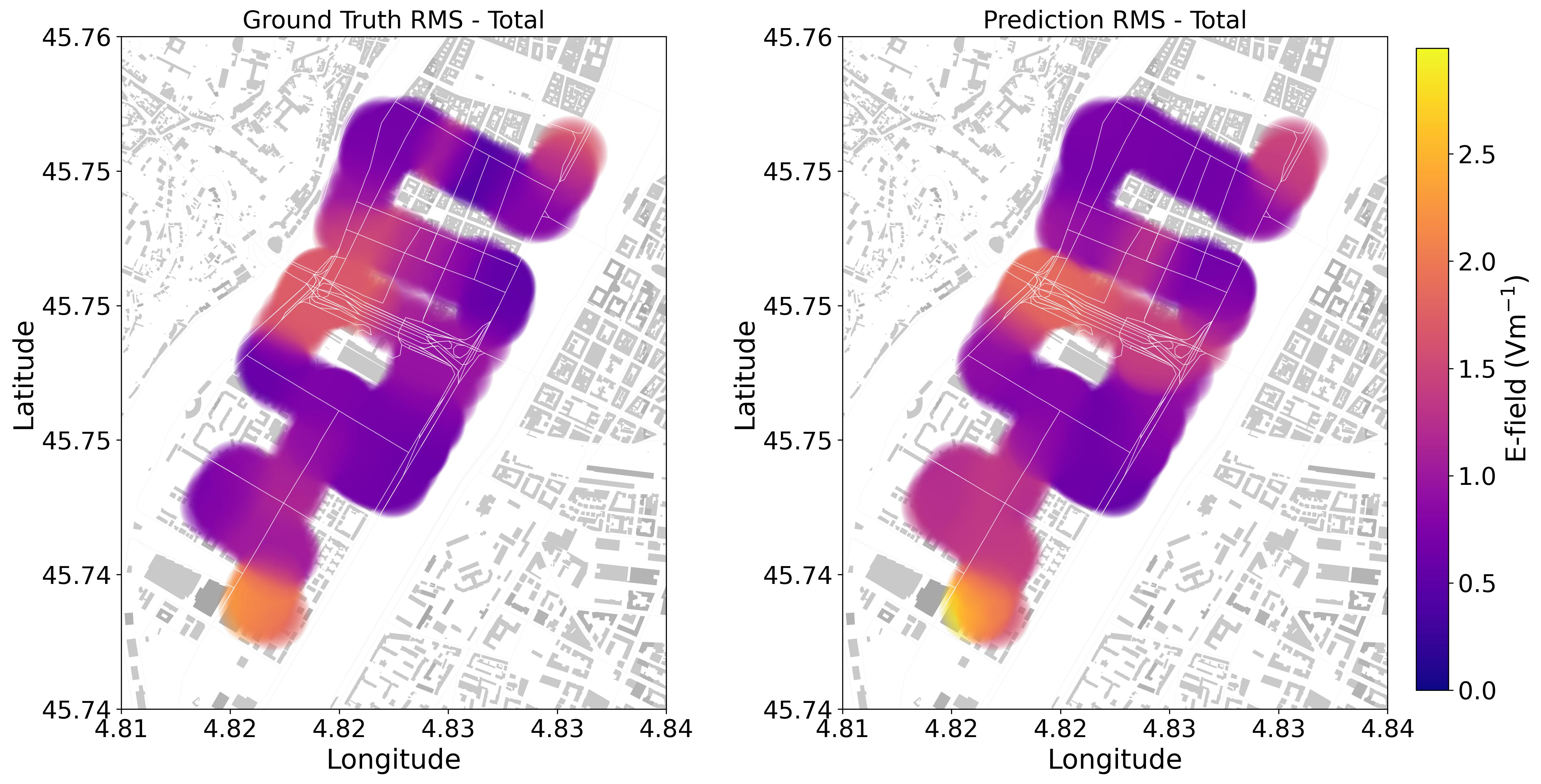}
    \caption{Results visualization and comparison for for testing dataset $2$ from Lyon.}
    \label{fig:MappingResults4}
\end{figure}

As shown in Table~\ref{tab:4}, both RMSE and MAPE values remain well below $27\%$, demonstrating a reliable predictive performance. Additionally, Fig.~\ref{fig:MappingResults4} shows that the spatial distribution of E-field levels is effectively captured, further validating the model's ability to generalize to different urban environments.

\section{Conclusions }\label{Section7}
In this study, a deep learning framework was developed to predict E-field levels in complex urban environments. A comprehensive dataset was constructed by integrating real-world drive test measurements, BSAs information, and geographic data.  Furthermore, unlike previous studies that extract features from the propagation environment and convert them into scalar values, this work directly represents these features as $2$-D input tensors. This approach not only makes them suitable for CNN-based models but also allows the underlying factors influencing EMF exposure levels, e.g. building layout and BSAs density, to be directly expressed in the $2$-D images, avoiding the potential information loss that occurs with scalar-based input representations.

Different from previous studies \cite{telecom_Wang}\cite{ChikhaAccess} where only the total E-field was predicted, the proposed model ExposNet was designed with two prediction modes: frequency-selective prediction and total E-field prediction, enabling adaptability to different types of datasets. The proposed network architecture was carefully designed to extract and integrate multi-modal environmental and antenna-related features while preserving their spatial relationships. It is worth noting that the output of ExposNet differs from previous studies, where neural networks were designed to predict the E-field value at each individual measurement point. Although a moving average method was applied to measurements within a chosen distance to filter out noise, consecutive measurement points still exhibited significant overlap. This redundancy resulted in too much similarities between training samples, limiting the generalization ability of the trained model.
In this study, rather than predicting the E-field level at each measurement point, we predict the average E-field value within a defined area, along with its STD. To reduce overlap, target areas were carefully selected with a certain separation distance. This strategy addresses the data redundancy issues found in previous approaches, leading to a more robust and generalizable model.

Extensive experiments demonstrated that ExposNet achieves good prediction accuracy, effectively capturing the spatial distribution and magnitude variations of E-field levels across multiple frequency bands, given that it is trained and tested exclusively on real-world measurement data. Furthermore, the model’s generalization ability was assessed using different testing dataset. First, the model was trained using a combined dataset of measurements from Paris and Lyon and evaluated on measurements from Paris. Then, it was tested on two different segments of Lyon’s measurement data, with training performed on the corresponding remaining dataset from Paris and Lyon. The results indicate that ExposNet maintains good predictive performance through different testing datasets. 

Future work will focus on further improving model generalization across diverse environments and extending the study on uncertainty quantification to enhance the model’s reliability and interpretability.

\section*{Acknowledgment}
This work was supported in part by EU Project SEAWAVE through the Horizon Europe Research and Innovation Program under Grant $101057622$.

\bibliographystyle{IEEEtran} 




\end{document}